\documentclass[acmsmall]{acmart}
\usepackage{tabularx}
\usepackage{enumitem} 
\usepackage[most]{tcolorbox}
\usepackage[htt]{hyphenat}

\newtcolorbox{promptbox}[2][]{
  colback=gray!5,      % Background color of the box
  colframe=gray!60,    % Frame color
  fonttitle=\bfseries, % Bold title
  title=#2,            % The title of the box
  arc=2mm,             % Rounded corners
  boxrule=0.5pt,       % Thickness of the frame
  #1                   % Allows for additional custom options
}

\usepackage{xcolor}
\definecolor{blue}{RGB}{0, 114, 189}
\definecolor{red}{RGB}{217, 83, 25}

\AtBeginDocument{%
  }

\begin{document}

\title[SpeakSoftly]{SpeakSoftly: Scaffolding Nonviolent Communication in Intimate Relationships through LLM-Powered Just-In-Time Interventions}

\author{Ka I Chan}
\affiliation{
  \institution{School of Information, University of Michigan}
  \city{Ann Arbor}
  \country{United States}}
\email{chankai@umich.edu}
\orcid{0009-0001-4560-1702}

\author{Hongbo Lan}
\affiliation{
  \institution{University of Pittsburgh}
  \city{Pittsburgh}
  \country{United States}}
\email{hol101@pitt.edu}
\orcid{0009-0003-7333-1863}

\author{Jun Fang}
\affiliation{
  \institution{Department of Computer Science and Technology, Tsinghua University}
  \country{China}}
\email{fangy23@mails.tsinghua.edu.cn}
\orcid{0009-0001-2614-8674}

\author{Yuntao Wang}
\affiliation{
  \institution{Key Laboratory of Pervasive Computing, Ministry of Education, Department of Computer Science and Technology, Tsinghua University}
  \country{China}}
\email{yuntaowang@tsinghua.edu.cn}
\orcid{0000-0002-4249-8893}

\author{Yuanchun Shi}
\affiliation{
  \institution{Key Laboratory of Pervasive Computing, Ministry of Education, Department of Computer Science and Technology, Tsinghua University}
  \country{China}}
\affiliation{
  \institution{Intelligent Computing and Application Laboratory of Qinghai Province, Qinghai University}
  \country{China}}
\email{shiyc@tsinghua.edu.cn}
\orcid{0000-0003-2273-6927}

\renewcommand{\shortauthors}{Chan et al.}

\begin{abstract}
Conflicts are common in text-based communication, particularly in intimate relationships, where misunderstandings can easily escalate into verbal aggression. To address this, we present \textit{SpeakSoftly}, a system that applies Nonviolent Communication (NVC) principles to scaffold couples' conflict communication through LLM-powered just-in-time interventions. Informed by formative interviews with couples and NVC principles, we designed two core features: NVC-Prompt, which detects verbal aggression and suggests revisions to prevent escalation, and NVC-Guide, which analyzes dialogues to uncover users' feelings and needs, fostering self-awareness and perspective-taking. These features were implemented across three progressive intervention modes, each varying in intervention depth and tone: Basic Reminder, Neutral Guide, and Empathetic Guide. We conducted a mixed-methods user study with 18 couples across simulated and real-life conflict settings to evaluate the effectiveness of each mode. Results showed that Empathetic Guide significantly facilitated both behavioral and cognitive changes, while Neutral Guide was effective only for behavioral changes in simulated conflicts. In real-life conflicts, Neutral Guide showed distinct advantages due to lower cognitive load demands. We discuss the mechanisms behind these findings and propose design implications for in-situ interventions in high-stakes communication contexts.
\end{abstract}

%% The code below is generated by the tool at http://dl.acm.org/ccs.cfm.



\begin{teaserfigure}
  \includegraphics[width=\textwidth]{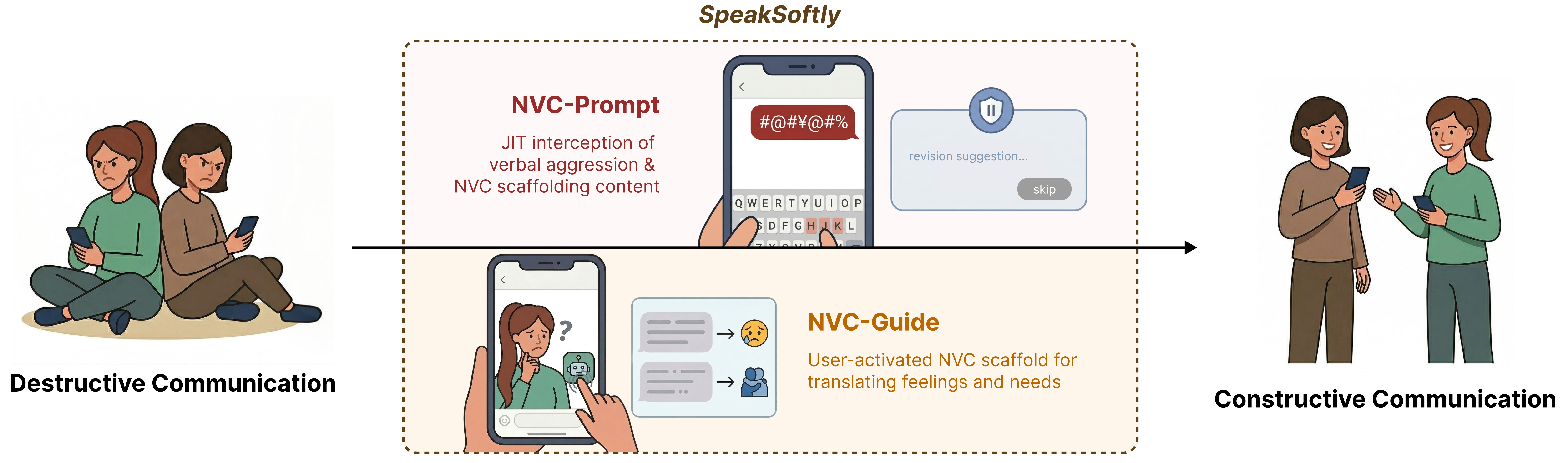}
  \caption{\textbf{Overview of \textit{SpeakSoftly}.} During text-based conflicts, the system provides two core features: NVC-Prompt, which intercepts verbal aggression with just-in-time revision suggestions, and NVC-Guide, which helps users translate mutual feelings and needs. Together, these features scaffold users toward constructive communication.}
  \Description{This figure illustrates the overall workflow of SpeakSoftly. On the left, two people sit back to back looking at their phones with frustrated expressions, representing destructive communication. In the center, enclosed in a dashed border labeled SpeakSoftly, two features are shown: NVC-Prompt at the top displays a phone screen with an intercepted aggressive message and a pop-up offering revision suggestions with a skip button; NVC-Guide at the bottom shows a phone screen analyzing conversation content with arrows pointing to emoji representing feelings and needs. On the right, the same two people stand facing each other with relaxed expressions, representing constructive communication. A horizontal arrow connects the left, center, and right sections, indicating the transformation from destructive to constructive communication through the system's intervention.}
  \label{fig:teaser}
\end{teaserfigure}

\maketitle

\section{Introduction}
\label{sec:intro}

% 冲突的不可避免性 -> 对于couple来说冲突的影响 -> communication在冲突中的重要性
Conflicts are inevitable, especially in intimate relationships, as no two individuals share identical perspectives, needs, or expectations. While conflicts can lead to negative emotional experiences, they can also promote personal growth and deepen relational bonds \cite{ames1982two, hendrick2000close}. Critically, relationship outcomes are shaped not by whether couples experience conflict, but by how they communicate during the conflict \cite{gottman2007marriages, markman1993preventing, winstok2012partner}. 
This challenge is particularly pronounced for couples who rely on text-based communication, where the absence of vocal tone and nonverbal cues makes misunderstandings more likely to occur and escalate \cite{johnson2016computer, kelly2018perceived}. Miscommunication in such contexts often manifests as verbal aggression \cite{johnston1994high}, which can trigger cycles of escalating hostility between partners \cite{sowan2024conflict}. These dynamics carry serious consequences, including links to adverse health outcomes \cite{eaker2007marital}. Addressing destructive communication patterns, such as insults, threats, and accusations, is therefore essential to mitigate harm and promote healthier conflict resolution \cite{gottman2007marriages}. 

% 引入沟通技巧的运用 -> NVC
Given the critical role of communication in conflict outcomes, adopting communication techniques can significantly improve relationship satisfaction and stability \cite{gottman2007marriages}. Therapists often emphasize the value of effective communication techniques in strengthening relationships \cite{williamson2016effects}. One widely recognized approach is Nonviolent Communication (NVC), developed by clinical psychologist Marshall Rosenberg \cite{rosenberg2015nonviolent}. Grounded in principles of nonviolence and humanistic psychology, NVC is structured around four key components: observation, feelings, needs, and requests, which is designed to enhance understanding, connection, and empathy between individuals. By guiding individuals to articulate their emotions and needs without blame, NVC empowers couples to transform conflicts into opportunities for growth and deeper emotional connection. 

% 目前HCI的亲密关系干预研究
Within the HCI community, researchers have explored technology-mediated interventions to facilitate connections \cite{bales2011couplevibe, griggio2019augmenting, yang2017communicating}, simulate physical presence \cite{park2012couples, park2013roles}, and enhance emotional expression \cite{chien2013whisper, li2018review} for couples. More recently, conflict mediation has received growing attention. For example, TogetherReflect provides art therapy for couples' conflict through VR \cite{VR-LDR}, PocketBot offers chatbot-based support to reconnect couples after conflicts \cite{pocketbot}, and LLM-powered systems enable users to rehearse predefined conflict scenarios or reflect on past conflicts \cite{chun2025conflictlens, rehearsal}. 
However, these existing tools focus primarily on pre-conflict preparation or post-conflict reflection, leaving a critical gap: the moment when conflict is actively unfolding remains largely unsupported, especially when destructive communication behaviors occur. Moreover, deploying chatbots as explicit third-party mediators raises concerns about fostering over-reliance that diminishes couples' capacity for independent resolution \cite{adebayo2024role}, and about being perceived as referees whose judgments may intensify competition and further escalate conflict \cite{a2000social, rubin1985third}.

% 本研究内容介绍
To address these gaps, we present \textit{SpeakSoftly}, a system that scaffolds couples' communication during active conflicts by providing large language model (LLM)-powered in-situ interventions grounded in NVC principles (see Figure~\ref{fig:teaser} for an overview). We first conducted formative interviews with nine experienced participants to identify the challenges couples face when attempting to apply communication techniques during conflicts. Informed by these findings and prior literature, we synthesized four design goals and developed \textit{SpeakSoftly} as a chat application incorporating two core features: NVC-Prompt, which is triggered automatically upon detecting verbal aggression and offers a temporary interception with revision suggestions; and NVC-Guide, which users activate to receive guidance in understanding their own feelings and needs and in empathizing with their partner's perspective. To examine the appropriate level of intervention depth, we implemented these features across three progressive modes: Basic Reminder, Neutral Guide, and Empathetic Guide. These modes vary in system involvement, tone, and scaffolding depth. We evaluated \textit{SpeakSoftly} through a within-subjects mixed-methods study with 18 couples across both simulated and real-life conflict settings. Our results demonstrate that Empathetic Guide was the most effective mode in simulated conflicts, producing significant behavioral and cognitive changes, while Neutral Guide showed distinct advantages in real-life conflicts where cognitive resources were more constrained.

To summarize, this paper makes the following contributions:
\begin{itemize}
    \item Empirical formative findings identifying the critical challenges couples face when applying communication techniques during conflicts, derived from semi-structured interviews.
    \item The design and implementation of \textit{SpeakSoftly}, an open-source system featuring NVC-Prompt and NVC-Guide that scaffolds couples' conflict communication through LLM-powered in-situ interventions grounded in NVC principles.
    \item Evaluation insights from a mixed-methods study demonstrating that empathetic scaffolding is the most effective intervention mode for fostering behavioral and cognitive change, with distinct patterns emerging between simulated and real-life conflict settings. 
    \item Design implications for in-situ interventions in high-stakes communication, including the value of deliberate communicative friction, scaffolding over automation, and adaptive intervention strategies calibrated to emotional intensity. 
\end{itemize}

\section{Background and Related Work}

\subsection{Theoretical Foundation: Nonviolent Communication and Destructive Communication Patterns}
\label{sec:nvc}
% NVC 的核心机制：简述观察、感受、需求、请求四个核心要素，并强调 NVC 在处理愤怒和亲密关系冲突中的实证有效性。
Nonviolent Communication (NVC), developed by Marshall Rosenberg, is a human-centered communication process designed to foster compassionate connection and mutual understanding \cite{rosenberg2015nonviolent}. The essence of NVC lies in the conscious application of four core components: observations (describing facts without evaluation), feelings (identifying specific emotions), needs (uncovering individual's requirements driving those feelings), and requests (proposing clear, specific, and positive actions to fulfill individual's needs without implying blame or punishment for non-compliance) \cite{rosenberg2015nonviolent}. In the context of intimate relationships, NVC is particularly effective because it reframes anger not as a destructive emotion, but as a crucial signal stemming from unmet needs combined with judgmental thoughts \cite{rosenberg2004we, little2008total}. Empirical studies, such as NVC-based communication programs for couples, have validated that applying these principles significantly increases empathy, improves conflict resolution skills, and enhances marital satisfaction \cite{nvc-couple, CA-couple, little2008total}. 

% 引用心理学文献说明：为什么情侣在吵架时，即便知道 NVC 也用不出来？因为人在高压下会自动退回到防御性的负面沟通习惯中。
However, couples often struggle to apply communication techniques in the heat of a real-life conflict \cite{ronan2004violent, fincham2003communication, fowers2001limits}. This challenge primarily stems from deeply entrenched, destructive communication tendencies. As conflicts emerge, unachieved goals and sensitivity to perceived hurt frequently trigger aggressive reactions \cite{sowan2024conflict}, including criticism, contempt, defensiveness, and stonewalling \cite{gottman2007marriages}. Consequently, the progression of conflict is marked by a steady decline in communication quality \cite{weingarten1987levels}. Instead of constructive dialogue, couples instinctively revert to dysfunctional methods of conflict resolution (e.g., competition and avoidance) \cite{weingarten1987levels} and distorted communication strategies (e.g., accusation, rationalization, and victimization) to favor their own perspectives \cite{whiting2016escalating}. Furthermore, this dynamic is highly reciprocal: the use of verbal aggression by one partner is positively correlated with reciprocal aggression from the other \cite{sowan2024conflict}. As these negative conflict behaviors continuously heighten emotional arousal \cite{gottman2023predicts}, couples become trapped in a vicious cycle. While effective conflict management theoretically relies on self-initiated repair attempts to reduce negative affect and increase positive affect \cite{gottman2015repair}, relying on individuals to consciously execute these repairs during intense emotional distress is often unrealistic. 

Because these destructive communication tendencies are often driven by impulsive and automatic thinking, merely learning communication techniques through prior education is frequently insufficient for in-situ application. Therefore, instead of retrospective assistance, a just-in-time (JIT) scaffolding is needed to effectively break the vicious cycle of verbal aggression and foster genuine empathy.

\subsection{Technology-Mediated Interventions for Intimate Relationships}
% 回顾现有 HCI 研究
Computer-mediated communication (CMC, e.g., text, voice, video) plays an increasingly important role in the initiation, development, and maintenance of interpersonal relationships, particularly romantic relationships \cite{rabby2003computer, tong12011relational, coyne2011luv}. Long-distance relationships (LDRs), in particular, heavily rely on these technological mediums to maintain intimacy and harmony \cite{neustaedter2012intimacy}. Among these channels, text-based communication remains the most frequently used \cite{hampton2017channels}. While asynchronous CMC allows users more time to craft purposeful messages \cite{walther2007selective, perry2011couples}, it is notoriously prone to misunderstandings due to the absence of physical touch, nonverbal and social cues \cite{sproull1986reducing, culnan1987information}. The reduction of these cues often leads recipients to overinterpret ambiguous language \cite{hancock2001impression}. When coupled with negative emotional states, individuals tend to interpret messages through the lens of their feelings rather than objective meaning, easily triggering conflicts \cite{walther1996computer, sitorus2025language}. For instance, \citet{scissors2016bias} shows that individuals with lower self-esteem are more susceptible to having negative biases triggered in instant messaging compared to face-to-face interactions, resulting in significantly worse assessments of conflict discussions. Consequently, text-based CMC is frequently cited as a catalyst for poor communication and is often deemed unsuitable for serious conversations, as it can rapidly escalate disputes \cite{kelly2018perceived, scissors2013back, kelly2012s}.

To mitigate these communication challenges and maintain intimacy, the HCI community has actively explored couple technologies, particularly for LDRs. For instance, contextual information streams allow partners to share daily routines (e.g., phone battery, steps, location) through smartphones, fostering a sense of connection without the need for proactive messaging \cite{griggio2019augmenting, bales2011couplevibe}. Other modalities aim to simulate physical presence; CheekTouch enables couples to send phone vibrations to a partner’s cheek during calls \cite{park2012couples, park2013roles}, while Whisper Pillow supports asynchronous audio messages to enhance emotional expression for partners with different schedules \cite{chien2013whisper}. While these technologies successfully foster emotional connection in everyday scenarios, they primarily focus on intimacy maintenance rather than addressing conflicts. As intimate relationships inevitably encounter disagreements, the research focus has increasingly expanded toward active conflict mediation.

Recently, various technologies have been utilized to facilitate conflict resolution. For example, TogetherReflect provides a collaborative VR experience based on art therapy, allowing couples to draw and visualize their emotions after a conflict to open new channels for mutual understanding \cite{VR-LDR}. For text-based CMC, chatbots like PocketBot offer features to help LDR partners reconnect after conflicts \cite{pocketbot}. Furthermore, some systems utilize large language models (LLMs) to create training sandboxes (e.g., Rehearsal, ConflictLens), allowing users to practice predefined conflict scenarios in advance and supporting self-guided reflection \cite{rehearsal, chun2025conflictlens}. Other approaches introduce virtual agents as a third-party integrated into smartphones to encourage communication when couples are in joint scenes \cite{khatra2024agent}. While their methods vary, these tools all share the goal of improving communication techniques and guiding couples toward more satisfying resolutions. 

% Gap ：(1) 事后反思/事前模拟，缺乏in-situ application；(2) 许多现有的聊天机器人充当了第三方裁判 (referees)，容易让情侣产生技术依赖，而不是真正教导他们沟通技巧。
Despite these advancements, existing interventions present notable limitations when addressing the deeply ingrained, destructive communication patterns discussed in Section~\ref{sec:nvc}. \citet{baughan2024supporting} highlights that technology should empower users to navigate emotional vulnerability rather than solving the problem for them in terms of hard conversation in close relationships. However, most current systems focus heavily on pre-conflict rehearsal or post-conflict reflection. As established, individuals often lose the cognitive capacity to apply learned skills during the intense emotional arousal of a real-time conflict, rendering these practices insufficient for in-situ behavioral change.

Moreover, utilizing chatbots as active third-party mediators poses significant risks. Relying heavily on external agents can establish technological dependency, reducing an individual's self-sufficiency in resolving conflicts independently \cite{adebayo2024role}. Worse, when a chatbot is perceived as an authoritative "referee," it can inadvertently trigger couples to compete over who is right or wrong, thereby exacerbating the very verbal aggression the system aims to reduce \cite{rubin1985third, a2000social}. To address these gaps, there is an urgent need for a paradigm shift: moving from offline sandboxes and authoritative referees toward an in-situ scaffolding approach. Such an ideal intervention must be capable of gently interrupting destructive communication patterns exactly when emotional arousal peaks, without usurping the users' autonomy in resolving their own conflicts.

\subsection{AI-Mediated Communication (AIMC)}
% 解释为什么 LLM 特别适合做这件事
With the rapid advancement of Large Language Models (LLMs), Artificial Intelligence-Mediated Communication (AIMC) has become increasingly prevalent beyond traditional computer-mediated communication. \citet{hancock2020ai} define AIMC as "mediated communication between people in which a computational agent operates on behalf of a communicator by modifying, augmenting, or generating messages to accomplish communication or interpersonal goals". Enhanced by capabilities such as semantic understanding and context reasoning \cite{naveed2025comprehensive}, LLMs can effectively identify verbal aggression in complex conflict scenarios. Furthermore, their context-aware and empathetic capabilities enable them to move beyond functional assistance to provide emotional support \cite{siddals2024happened}. Because LLMs offer a non-judgmental, time-unrestricted, and safe environment, individuals are often more willing to self-disclose sensitive topics \cite{lee2020designing, zheng2025customizing}, thereby demonstrating the potential to uncover users' underlying unmet needs during conflicts.

% 高风险场景下的痛点
Despite the proliferation of AIMC tools designed to facilitate communication, researchers caution against over-relying on AI when mediating relationships. Specifically, when AI acts as a full mediator or generates automatic responses, it creates "Empathy Fog": a state of uncertainty where it becomes difficult to identify how much genuine empathy, attention, and effort the human sender actually invested in the conversation \cite{wolfe2025toward}. This concern is echoed by users' perceptions. \citet{fu2024text} categorized the impact of AIMC tools and found them particularly relevant for high-stakes communication scenarios, such as arguing with friends or romantic partners. While users appreciate AIMC tools for building communication confidence and finding accurate expressions, they strongly oppose AI automatically responding on their behalf. Instead, they prefer the AI to act as a "wise friend" or a "communication coach" that offers reflective suggestions while strictly preserving user autonomy.

To combat Empathy Fog and ensure AI facilitates rather than replaces human connection, \citet{wolfe2025toward} proposed Needs-Conscious Design, a human-centered framework for AIMC grounded in NVC principles. This framework is structured around three core pillars: Intentionality (slowing down fast-paced digital communication to foster self-reflection), Presence (ensuring human effort and authenticity are not obscured by AI), and Receptiveness to Needs (shifting the focus from determining right or wrong to uncovering unmet needs). However, it remains largely unexplored how these macro-level design pillars translate to the highly sensitive, emotionally volatile context of romantic text-based conflicts. As our formative study reveals, informal, high-stakes interactions often involve intense emotions where communication techniques frequently fail. Therefore, an ideal AIMC tool in this specific context should function as a reflective coach that regulates emotions and guides users toward these three pillars, rather than an autonomous agent that takes over the conversation.

To bridge this critical contextual gap and operationalize the Needs-Conscious paradigm within intimate partner dynamics, our study introduces \textit{SpeakSoftly}, an in-situ scaffolding system driven by LLMs and grounded in NVC principles. Rather than acting as a referee or an offline sandbox, our system provides a JIT intervention to gently interrupt destructive communication patterns exactly when emotional arousal peaks, empowering couples to explore the underlying needs driving the conflict and cultivate mutual empathy even in high-stakes scenarios. 
\section{Formative Study}

We sought to improve our understanding of the experience and challenges couples face when attempting to apply communication techniques in real-world conflict scenarios, we conducted a exploratory formative study through semi-structured interviews. 

\subsection{Methods}
Participants were recruited through an online advertisement distributed on social platforms. Eligible participants were required to have prior experience using communication techniques (such as NVC) during conflicts with current or past romantic partners, and to regularly use text-based CMC for conflicts. We excluded individuals who failed to complete the screening questionnaire thoroughly. Ultimately, three couples and three individual participants (9 participants in total; 5 females and 4 males, age ranged from 21 to 35, \textit{M} = 26.33, \textit{SD} = 5.32) were recruited. They represented diverse backgrounds in relationship length and distance settings. To maintain anonymity, couples were assigned letters A through C (e.g., A1 and A2) while individual participants were labeled D through F (e.g., D). The study was approved by our Institutional Review Board (IRB), and participants were fully informed about the study's purpose and received monetary compensation after data collection. 

All interviews were conducted remotely via videoconferencing and audio-recorded with participants' consent. Each interview lasted between 40 to 75 minutes. The interviews aimed to explore conflict experiences in romantic relationships and the use of communication techniques. The first part focused on conflict experiences, with questions like: \textit{"What were your conflict scenarios like?"}, \textit{"What escalated the conflict?"}, \textit{"Have you experienced verbal aggression? If so, what was it like?"}, and \textit{"How do you resolve conflicts eventually?"} The second part addressed communication techniques, with questions like: \textit{"How did you use [specific communication techniques]?"}, \textit{"Do you consider your experience in using this technique for conflicts as a success or failure?"}, and \textit{"What challenges did you encounter when attempting to use communication techniques?"}. The recordings were transcribed by the first author who conducted all interviews. The first author coded the transcripts to identify emerging themes. The research team then engaged in several discussions to refine and consolidate these themes into key findings presented below.

\subsection{Findings}
\label{sec:formative}

From the interviews, we identified two primary challenges that illustrate how our participants' conflicts frequently degraded from collaborative discussions into destructive arguments, ultimately obstructing their ability to effectively apply communication techniques. 

\subsubsection{Challenge 1: Emotional hijacking overrides learned communication techniques}
Participants described that their conflicts often originated from minor misunderstandings or differing viewpoints. However, when neither partner yielded, negative emotions quickly accumulated, causing them to misinterpret each other's words as blame or criticism. They reported that this rapid emotional escalation overwhelmed their self-control, rendering them unable to employ any learned communication techniques and frequently leading to instantaneous verbal aggression. As D described, conflicts often start small but spiral out of control: \textit{"Initially, I was just a little angry, but since we both stuck to our own opinions, the more we talked, the angrier we got... eventually leading to swear words and personal attacks."} Furthermore, E and F admitted to using destructive language to vent their emotions during heated moments, employing exaggerated threats like \textit{"let's just break up"}, dismissive remarks such as \textit{"whatever"} and \textit{"let it be"}, or even personal attacks like \textit{"Your temper is so awful; who would want to be in a relationship with you?"} Similarly, C1 noted that when feeling extremely frustrated, impulsive reactions often take over, leading to highly unpleasant exchanges and resorting to verbal aggression.

Consequently, this emotional hijacking frequently caused our participants to completely lose focus on the objective problem, transforming what could have been a collaborative conversation into a destructive one. As D claimed, the sheer difficulty of maintaining self-control during these moments causes a complete inability to utilize communication techniques effectively. A1 insightfully summarized this phenomenon: \textit{"All I remember are the emotions from the argument, but I’ve forgotten many of the reasons behind it."} This state of being overwhelmed by automatic emotional reactions provides a concrete, context-specific illustration of the severe lack of \textit{Intentionality} in natural conflict dynamics, a concept emphasized by \citet{wolfe2025toward}. When users are driven by reflexive anger rather than conscious choice, communication techniques fail at the very first step. This underscores the critical need for an in-situ, JIT intervention to enforce a brief, reflective pause before they speak.

\subsubsection{Challenge 2: Inertia of destructive communication patterns}
Even when emotions are somewhat regulated, participants reported that effective collaboration remains elusive due to the powerful inertia of entrenched destructive communication patterns.

\textbf{Challenge 2a: Fixation on the "right vs. wrong" dynamics triggers defensiveness.}
A recurring theme among our participants was the tendency to treat conflicts as adversarial battles, where partners viewed each other as the party at fault. A1 noted that her partner often perceived her expression of feelings as criticism, prompting him to become defensive. Describing a conflict regarding a trip to Disneyland, A1 shared how her partner (A2) misinterpreted her frustration over the itinerary as blame: \textit{"I do express my feelings or thoughts during arguments, but it seems like I unintentionally make him take the situation as me blaming him. Even though my intention might be to discuss the process, he sees it as me presenting evidence to prove he's wrong."} This defensive mindset of viewing a partner as an adversary severely prevents collaborative problem-solving. Similarly, E documented how she and her partner would fixate on identifying the "wrongdoer," even establishing reward-punishment rules for their conflicts. Furthermore, F claimed that it was difficult to understand why his partner was angry when he felt he "did nothing wrong," highlighting a severe blockage in empathy caused by this right-versus-wrong paradigm.

\textbf{Challenge 2b: Fixation on surface-level blame obscures underlying needs.}
Consequently, this adversarial framing often led participants to focus solely on surface-level behaviors, missing or misinterpreting their partner's underlying emotional needs. This pattern frequently manifested as a misalignment between behavioral fixes and emotional validation. For example, during a recurring dispute over late-night habits and drinking milk tea, B1 viewed the situation as a simple behavioral issue to correct, citing his own healthy routine. Conversely, B2 felt her underlying needs for autonomy and emotional understanding were fundamentally ignored: \textit{"He often says things like, 'Why do you eat such spicy food? Your acne hasn't even cleared up yet.' Then he starts explaining the reasons behind it... Sometimes, I do listen to him, and in the moment, it feels like the issue is resolved. But honestly, this habit is really hard to change. Later, I’ll bring it up again and try to negotiate with him."} Similarly, E described how her partner's reaction to forgetting their anniversary exacerbated the conflict. Instead of recognizing her need for emotional reassurance, he responded defensively: \textit{"Instead of apologizing, he said, ‘I’m not someone who checks the calendar often...’ This made me furious, and I replied, ‘Does this anniversary only exist because of a calendar?’ He dismissed it and kept making excuses."} To force her partner to acknowledge the seriousness of her unmet needs, E admitted to habitually escalating the situation by threatening a breakup: \textit{"After I propose breaking up, he suddenly seems stunned and realizes the seriousness of the issue. Then he keeps apologizing profusely and tries to make it up to me."} 

\textbf{Challenge 2c: Unrecognized emotional effort undermines sustained communication.}
Our participants highlighted that applying communication techniques during conflicts requires immense emotional self-regulation, which, if unreciprocated, can easily devolve into a feeling of swallowing grievances. For instance, A2 shared his struggles with attempting to proactively de-escalate arguments: \textit{"I try to make concessions... to ease the tension or apologize first, and then review the issue later. But sometimes I genuinely feel it is not my fault, so I cannot proactively de-escalate every single time."} Moreover, he emphasized that restraining his emotions through rational communication techniques often felt like an unacknowledged sacrifice, ultimately discouraging further effort: \textit{"I try to use a seemingly rational method to restrain my emotional loss of control, but it just makes me feel wronged. Afterwards, I don't feel like I received any reward or benefit for using this approach, nor did it make things more effective, so I simply abandon further attempts."}

Overall, these entrenched interaction patterns directly echo the lack of \textit{Receptiveness to Needs} outlined by \citet{wolfe2025toward}. When participants became fixated on determining fault or correcting surface behaviors, they lost the capacity to uncover the unmet emotional needs driving the conflict. Furthermore, without positive reinforcement or mutual acknowledgment for their constructive behaviors, participants reported quickly losing motivation and regressing into familiar defensive mechanisms. This combined challenge highlights the critical necessity for an in-situ AI mediator designed to empower users to explore and articulate their underlying needs, rather than fixating on behavioral blame. Additionally, it underscores the need for positive reinforcement mechanisms within the system to acknowledge users' emotional efforts and sustain constructive communication over time.

\section{SpeakSoftly: Design and Implementation}
\label{sec:design}
In this section, we present the design and implementation of \textit{SpeakSoftly}. Drawing upon the challenges identified in our formative study and prior literature across multiple disciplines, we first synthesize four design goals (Section~\ref{sec:dg}). We then translate these goals into specific system features (Section~\ref{sec:feature}) and bundle them into three intervention modes to be evaluated in our user study (Section~\ref{sec:mode}). Finally, we detail the system implementation through an LLM-powered web application (Section~\ref{sec:implementation}).

\begin{table*}[t]
\centering
\caption{Mapping formative findings and prior literature to our design goals (DGs) and the resulting system features in \textit{SpeakSoftly}.}
\label{tab:design_goals}
\renewcommand{\arraystretch}{1.4}
\small 
\begin{tabularx}{\textwidth}{
  >{\raggedright\arraybackslash}p{0.23\textwidth} 
  >{\raggedright\arraybackslash}p{0.20\textwidth} 
  >{\raggedright\arraybackslash}p{0.23\textwidth} 
  >{\raggedright\arraybackslash}X
}
\toprule
\textbf{Formative Findings} & \textbf{Prior Literature} & \textbf{Design Goals (DGs)} & \textbf{System Features \& Modes (Section~\ref{sec:feature} \& \ref{sec:mode})} \\ 
\midrule

% Row 1
\textbf{Challenge 1:} Emotional hijacking overrides learned communication techniques. & 
\textbf{Intentionality \& Conversational Friction} \newline \cite{wolfe2025toward, baughan2024supporting} & 
\textbf{DG1:} Interrupt impulsive reactions to prompt conscious reflection. & 
\textbf{NVC-Prompt:} Providing a JIT prompt to intercept messages containing verbal aggression. \\ 

% Row 2
\textbf{Challenge 2a:} Fixation on "right vs. wrong" dynamics triggers defensiveness. & 
\textbf{Referee Dynamics \& Empathy Fog} \newline \cite{adebayo2024role, a2000social, rubin1985third, wolfe2025toward} & 
\textbf{DG2:} Scaffold constructive expression without acting as a referee or obscuring human presence. & 
\textbf{NVC-Prompt \& NVC-Guide:} Asymmetrical UI; guiding expression instead of generating auto-replies. \\ 

% Row 3
\textbf{Challenge 2b:} Fixation on surface-level blame obscures underlying needs. & 
\textbf{Receptiveness to Needs} \newline \cite{wolfe2025toward, Rosenberg_2015} & 
\textbf{DG3:} Empower non-judgmental exploration of mutual needs. & 
\textbf{NVC-Guide:} User-activated function to foster independent understanding and translation of mutual needs. \\ 

% Row 4
\textbf{Challenge 2c:} Unrecognized emotional effort undermines sustained communication. & 
\textbf{Emotion Work \& \newline Positive Reinforcement} \newline \cite{erickson2005emotion, erickson1993reconceptualizing, holm2001association, skinner1963operant} & 
\textbf{DG4:} Cultivate constructive habits via positive reinforcement. & 
\textbf{Empathetic Guide:} Positive reinforcement mechanisms including thumbs-up emojis, a scoreboard, and tone-customized validation. \\ 

\bottomrule
\end{tabularx}
\end{table*}

\subsection{Design Goals}
\label{sec:dg}
Drawing upon the challenges identified in our formative study and prior literature across multiple disciplines, we synthesize four design goals (DGs) that \textit{SpeakSoftly} should meet. 

\textbf{DG1: Interrupt impulsive reactions to prompt conscious reflection.}   
Our formative study revealed that conflicts often escalate rapidly, hijacking users' emotional control and causing learned communication techniques to fail (Challenge~1). According to the Needs-Conscious Design framework, overcoming this destructive inertia requires cultivating intentionality. Specifically, slowing down to reflect on word choices and consciously valuing the needs of both partners, which is central to forming empathetic connections online \cite{wolfe2025toward}. Similarly, \citet{baughan2024supporting} proposes that increasing friction or providing reflection tips can break the habit of impulsive responses and make communication more effective for supporting hard conversations in close relationships. Therefore, the system should employ a JIT intervention mechanism that introduces a brief, gentle friction to prevent verbal aggression, enforcing a reflective pause exactly when emotional arousal peaks. 

\textbf{DG2: Scaffold constructive expression without acting as a referee or obscuring human presence.} 
During conflicts, participants frequently felt defensive when judged, making external interventions highly sensitive (Challenge~2a). Prior literature warns that introducing third-party interference into intimate conflicts often exacerbates hostility and negatively impacts relationship satisfaction, as it risks assuming the role of a referee \cite{rubin1985third, a2000social, adebayo2024role}. Conversely, if an AI simply takes over the conversation to avoid conflict, it risks creating Empathy Fog: a phenomenon where AI auto-generation obscures authentic human effort and presence \cite{wolfe2025toward}. To navigate these dual risks of AI intervention, the system should act as a private educational scaffold. By making all features visible only to the individual user (an asymmetrical interface), the system avoids acting as a shared referee. Furthermore, by only guiding users rather than sending automatic replies on their behalf, it preserves users' autonomy, authentic presence, and emotional effort in the conversation.

\textbf{DG3: Empower non-judgmental exploration of mutual needs.} 
As identified in Challenge~2b, participants often remain deadlocked because they fixate on surface-level behavioral blame while failing to interpret underlying emotional needs. According to NVC principles \cite{Rosenberg_2015}, human conflicts essentially arise from unmet needs, making the identification of the needs driving negative feelings a critical step for genuine resolution. Building on this, \citet{wolfe2025toward} emphasizes the concept of Receptiveness to Needs: the importance of shifting people's focus away from surface blame toward uncovering the unmet needs. Therefore, the system should provide a non-judgmental space to empower users to independently explore, translate, and articulate their own and their partner's needs, ultimately facilitating deeper self-reflection and empathy. 

\textbf{DG4: Cultivate constructive habits via positive reinforcement.} 
Applying communication techniques during heated conflicts requires immense emotional effort. Sociological literature defines this effort as emotion work, which involves attempts to "effectively manage the emotional climate within a relationship" \cite{erickson2005emotion}. As \citet{erickson2005emotion} notes, expressing empathy for a partner's feelings, especially when those feelings are not mutually shared, requires significant "time, effort, and skill" and represents "emotion work of the highest order". The critical role of this effort is evidenced by prior empirical studies, which demonstrate a negative association between emotion work and marital burnout \cite{erickson1993reconceptualizing}, and show that relationship satisfaction peaks when couples provide similar levels of emotion work \cite{holm2001association}. However, as highlighted in our formative findings, the lack of acknowledgment or positive feedback during intense conflicts often causes partners to feel wronged and abandon their constructive efforts (Challenge~2c). Therefore, drawing on behavioral principles of positive reinforcement \cite{skinner1963operant}, the system should incorporate positive reinforcement mechanisms. By acknowledging and validating users' emotional efforts, the system can help sustain motivation and gradually cultivate users' constructive communication habits.

\subsection{System Features}
\label{sec:feature}
To realize our design goals, we developed \textit{SpeakSoftly}, an LLM-powered chat application. The system intervenes during text-based conflicts based on NVC principles through two core features: \textbf{NVC-Prompt} and \textbf{NVC-Guide}. These features are encapsulated within an asymmetrical interface, meaning the interventions are visible only to the individual user, avoiding the risk of the system acting as a shared referee judging "right vs. wrong" between partners (\textbf{DG2}).

\textbf{NVC-Prompt: JIT interception for verbal aggression.}
NVC-Prompt operates as a JIT interception mechanism. By analyzing the message the user intends to send, the LLM determines whether it contains verbal aggression (e.g., personal judgments, labeling, commands, or accusations) aimed at their partner. If detected, the system temporarily intercepts the message and triggers a pop-up. This mechanism introduces a brief, intentional conversational friction that forces users to pause their impulsive emotional reactions. This pause creates an opportunity for emotional regulation, self-reflection, and reconsidering word choices before the damage is done (\textbf{DG1}). 
Furthermore, the triggered pop-up provides contextualized guidance based on the four core steps of NVC (observations, feelings, needs, requests) and suggests a revised framing of the message generated by the LLM. Crucially, the system does not auto-send the corrected message on the user's behalf. This scaffolds users to apply communication techniques and express themselves constructively, acting as a safeguard against irreversible verbal damage (\textbf{DG2}). To preserve individual autonomy and authentic presence, users are entirely free to accept, modify, or ignore the suggestions presented in the pop-up (e.g., by clicking a "skip" button) (\textbf{DG2}). 

\textbf{NVC-Guide: user-activated scaffold for needs translation.} 
NVC-Guide operates as a user-activated tool designed to foster an independent understanding of mutual feelings and needs during conflicts. When a user clicks the robot icon button, the LLM analyzes the recent chat history and provides a tailored analysis that quotes specific statements from the conversation. This analysis helps users empathize with their partner's emotions and translates surface-level complaints into underlying unmet needs. By functioning as a private educational scaffold, it reinforces the core tenet of NVC: understanding one's own and the other's feelings and needs without judgment (\textbf{DG3}).

\subsection{Intervention Modes}
\label{sec:mode}
\begin{figure}[b]
    \centering
    \includegraphics[width=\textwidth]{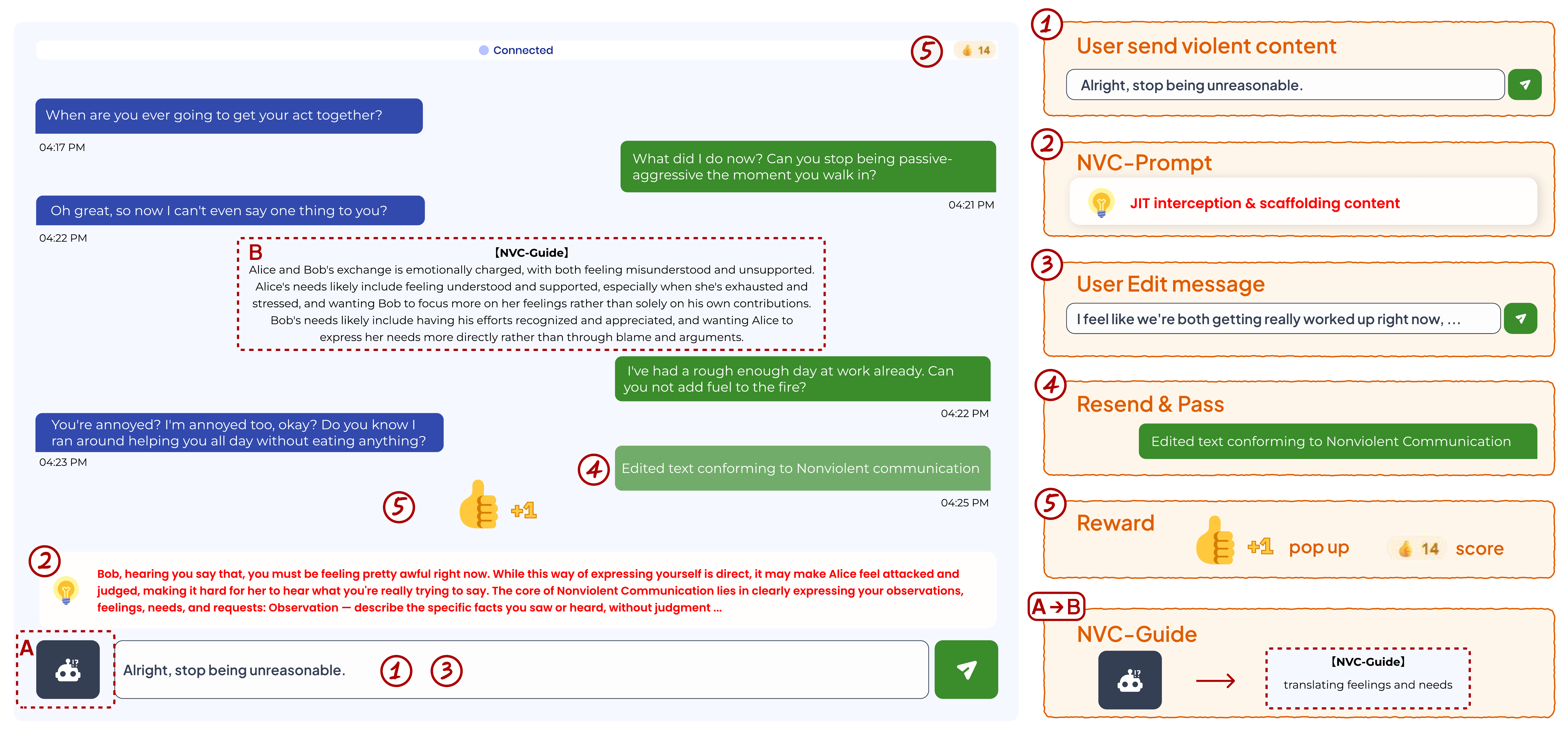}
    \caption{\textbf{System Interface of Empathetic Guide mode}, illustrated with a simulated conversation between two fictional users (Alice and Bob). The numbered steps indicate the interaction flow: (1) the user sends a message containing verbal aggression; (2) the system detects the aggression and triggers the NVC-Prompt, which provides JIT interception with scaffolding content (see Figure~\ref{fig:realexample} for detailed examples of NVC-Prompt suggestions across different modes); (3) the user edits their message based on the suggestion; (4) the revised message, now conforming to NVC principles, is sent to the partner; (5) the system rewards the user's effort with a thumbs-up pop-up and updates the score. (A->B) The NVC-Guide (bottom right) is a separate user-activated feature that analyzes the conversation to translate mutual feelings and needs.}
    \Description{This figure shows the system interface of SpeakSoftly in Empathetic Guide mode. The left side displays a chat interface with a simulated conversation between two users, Alice and Bob, who are having a conflict about household responsibilities. Numbered annotations on the interface mark five steps: (1) at the bottom, a text input field shows the user's original aggressive message "Alright, stop being unreasonable" with a send button; (2) a pop-up appears in the chat area containing the NVC-Prompt, which analyzes the user's emotions and suggests a revised message using NVC principles; (3) the input field shows the user's edited message; (4) a green chat bubble shows the revised message sent to the partner, marked as conforming to Nonviolent Communication; (5) a thumbs-up emoji with a plus-one indicator appears, and a score of 14 is displayed in the top right corner. The right side of the figure contains a step-by-step summary panel with numbered labels corresponding to the five steps on the chat interface. At the bottom right, a separate section labeled NVC-Guide shows a robot icon with an arrow pointing to a card that reads "translating feelings and needs," indicating the user-activated analysis feature. Two NVC-Guide outputs also appear within the chat as red-bordered messages labeled with B, providing analysis of the partner's feelings and needs.}
    \label{fig:system}
\end{figure}

To evaluate the optimal balance between intervention depth and emotional scaffolding, we implemented the NVC-Prompt and NVC-Guide through three progressive intervention modes. The three modes are designed to progressively layer our design goals, enabling us to empirically isolate the contribution of each design component. Basic Reminder implements only DG1, testing whether a JIT pause alone is sufficient to reduce verbal aggression. Neutral Guide adds DG2 and DG3, testing whether structured NVC scaffolding can shift users from surface-level blame toward exploring underlying feelings and needs. Empathetic Guide further adds DG4, testing whether emotional validation and positive reinforcement can sustain constructive efforts and foster deeper cognitive shifts. By comparing these three progressive modes, we can identify which design components are necessary and sufficient for behavioral change, cognitive shift, and sustained engagement. Users interact with these modes based on varying levels of system involvement and tone customization: 

\begin{itemize}
    \item[1.] \textbf{Basic Reminder:} A minimalist approach that triggers the NVC-Prompt strictly as a fixed reminder ("Verbal aggression detected") without providing specific revision suggestion or tone customization. This serves as a baseline for conversational friction.
    
    \item[2.] \textbf{Neutral Guide:} Integrates both the NVC-Prompt and NVC-Guide using a highly objective, robotic tone. It provides structured guidance (outlining observations, feelings, needs, and requests) but maintains emotional distance.
    
    \item[3.] \textbf{Empathetic Guide (with positive reinforcement mechanisms):} Building on the Neutral Guide, this mode supports the emotion work required to navigate conflicts. It adopts a friendly, highly empathetic persona to make the user feel cared for rather than lectured. Furthermore, it incorporates positive reinforcement mechanisms to sustain positive feedback loops: if a user successfully revises an intercepted message, the system rewards their emotional effort with thumbs-up emojis and points on a private scoreboard. This combination of intrinsic emotional validation and extrinsic gamification cultivates constructive communication habits (\textbf{DG4}). Figure~\ref{fig:system} illustrates the system interface of the Empathetic Guide mode. 
\end{itemize}

\subsection{System Implementation}
\label{sec:implementation}
We implemented \textit{SpeakSoftly} as a web application designed to facilitate text-based conflict discussions based on NVC principles. Upon reaching the login page, users can select from four system configurations: a baseline (no intervention) and the three intervention modes described in Section~\ref{sec:mode}. Users input a unique, self-determined username, which the AI uses to address them personally, and are then authenticated into a shared chat room with their partner where chat history is retained. 

% \textbf{System Architecture.} 
At the architectural level, the system's real-time communication functionality is powered by a Python-based backend using FastAPI, which serves both REST APIs and WebSocket connections for seamless message exchange. MongoDB is utilized as the database (hosted on a secure institutional laboratory server) to persistently store chat histories. The frontend chat interface is developed in JavaScript using React.js, handling both user-to-user WebSocket communication and user-to-LLM interactions. 

% \textbf{LLM Integration and JIT Interception Logic.} 
To enable the LLM integration, we leveraged the Google Gemini API (specifically, the \texttt{gemini-2.5-flash-lite} model) to power the core features. This model was selected for its optimal balance of performance, low response latency, and context window size. Additionally, the Gemini API's configurable safety filters were particularly crucial, as they allow the system to process and analyze destructive communication without triggering automatic service refusals.

To realize the JIT interception, the frontend temporarily holds the user's intended message and sends it to the FastAPI backend. The backend queries the LLM to classify the message based on predefined NVC criteria. Specifically, if the LLM detects verbal aggression, which defined in our instructions as personal judgments, labeling, commands, or accusations (see Appendix~\ref{apd:prompt}), the message is intercepted and the frontend renders the NVC-Prompt. If the message is deemed safe, or if the user chooses to skip the prompt, it is instantly emitted to the partner via the WebSocket.

% \textbf{Prompt Engineering.} 
In terms of prompt engineering, we developed distinct instruction prompts tailored to the specific persona and task for each intervention mode (detailed in Appendix~\ref{apd:prompt}). These prompts were formulated using chain-of-thought (CoT) reasoning and few-shot learning strategies to prevent the AI from overstepping. Specifically, to ensure the system does not act as a referee, the CoT prompt instructs the LLM to first internally evaluate the input category and strictly adhere to the "minimum necessary intervention" required before generating a response. To ensure the NVC-Guide provides highly contextualized advice, the backend maintains a sliding context window of the 20 most recent messages. Rather than merely avoiding token limits, this constrains the model's analytical scope to the most relevant conversational thread, ensuring that advice remains highly contextualized and responsive to the latest emotional shifts.

% \textbf{Data Privacy.} 
Finally, to ensure robust data privacy, given the sensitive nature of couples' conflict discussions, we implemented strict privacy safeguards. To ensure data sovereignty, participant data is pseudonymized via self-determined usernames. All chat histories are persistently stored on a secure institutional laboratory server. To leverage LLM capabilities while minimizing exposure, all interactions with the Gemini API are designed as stateless, single-turn requests. Under enterprise API terms, the data is processed ephemerally, ensuring that conversation data is neither retained by the service provider nor used for model training.

\section{User Study}
To evaluate the optimal design and effectiveness of the three intervention modes, we conducted a two-phase user study: a simulated conflicts phase in a controlled lab setting, and a real-life conflict phase in a natural field setting. We acknowledge the exploratory nature of this study.

\subsection{Participants}
\label{subsec:participants}
Assuming a medium effect size of 0.23 \cite{ortloff2025small}, an alpha level of 0.05, and a power of 0.8, a G*Power analysis suggested a required sample size of 28. We recruited participants through an online advertisement posted on social media, targeting couples who frequently experience text-based conflicts involving verbal aggression. All participants were screened to ensure they were at least 18 years old. Initially, 42 participants (21 couples) were recruited. Six participants (3 couples) withdrew during the experiment due to scheduling conflicts. An additional five participants were excluded because they failed to use a monitor of 10 inches or larger as required for the simulated conflict phase; in one case, only one member of a couple was excluded, resulting in an odd-numbered final sample. The final sample for the simulated conflict phase comprised 31 participants (15 females, 16 males; age: $M = 23.65$, $SD = 3.42$; relationship length: $M = 2.60$, $SD = 1.98$ years). Although this fell marginally below the target of 32, the shortfall of one participant has a negligible impact on statistical power. The cohort included 11 college students, 1 unemployed individual, and 19 individuals from various occupations. Among them, 14 participants (7 couples; 7 females, 7 males; age: $M = 25.00$, $SD = 4.10$; relationship length: $M = 3.50$, $SD = 2.51$ years) advanced to and completed the real-life conflict phase. To prioritize participants' privacy and emotional well-being during the real-life conflict phase, we implemented an additional ethical safeguard: couples were explicitly given the flexibility to bypass our system and use their accustomed communication channels for any conflicts they deemed too sensitive or private. This privacy-preserving design naturally resulted in anticipated data attrition during the field deployment. All participants provided informed consent and received monetary compensation for their participation.

\subsection{Study Design}
We employed a within-subject design with four intervention conditions as the main independent variable: a baseline (no intervention) and three intervention modes (Basic Reminder, Neutral Guide, and Empathetic Guide). The study spanned approximately 5 weeks, comprising an 8-day simulated conflict phase and a 4-week real-life conflict phase. Each couple used each intervention mode twice: once during the simulated conflict phase and once during the real-life conflict phase. To mitigate order effects, the sequence of intervention conditions was randomly assigned and counterbalanced, maintaining the identical sequence across both phases to ensure consistency in user experience. Furthermore, a one-day break was enforced between intervention conditions in the simulated conflict phase as a washout period to reduce the emotional impact and carryover effects of prior intervention conditions. The study was approved by our IRB.

\begin{figure}
    \centering
    \includegraphics[width=\textwidth]{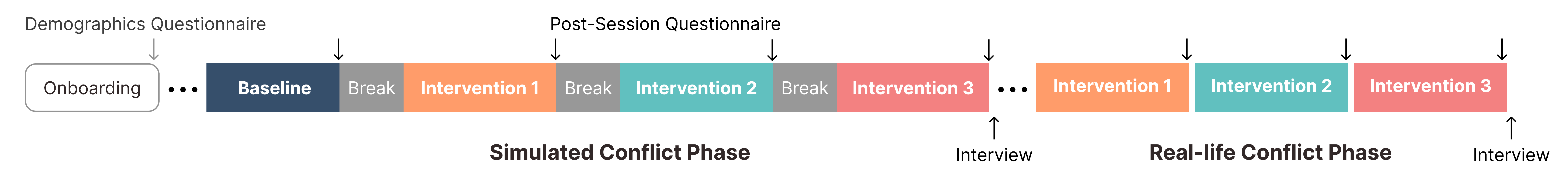}
    \caption{\textbf{Procedure of User Study.} The sequence of the three intervention modes was randomly assigned and counterbalanced. A one-day washout period was scheduled between each session in the simulated conflict phase to minimize the emotional impact and carryover effects of the previous session.}
    \label{fig:procedure}
    \Description{This is a flowchart showing the timeline of the user study, which consists of several phases. (1) Onboarding: Participants complete a demographics questionnaire. (2) Simulated Conflicts Phase: This phase includes a baseline session, followed by three randomized intervention modes, with a one-day washout period between each. A post-session questionnaire is completed after each session. (3) Exit Interview: A brief interview is conducted after the simulated phase. (4) Real-Life Conflicts Phase: Participants use the same sequence of the three intervention modes in a natural setting. A post-session questionnaire is completed after each conflict. (5) Final Exit Interview: The study concludes with a brief interview.}
\end{figure}

\subsection{Procedure}
Figure~\ref{fig:procedure} provides an overview of the user study procedure. 

\subsubsection{Onboarding}
Participants provided informed consent for their data to be used anonymously. Once enrolled, they completed a demographics questionnaire and specified a username that the AI would use to address them in the interventions.

\subsubsection{Simulated Conflicts Phase}
This phase consisted of four sessions (baseline plus three intervention modes) with a one-day break between each session, spanning a total of eight days. To ensure rigorous experimental control and address potential ethical concerns, we strictly calibrated the intensity of the conflict topics, referring to \citet{afifi2012standards} and \citet{scissors2016bias}. First, participants individually listed potential conflict topics, explicitly excluding any they considered too sensitive or did not wish to discuss. Subsequently, couples met with researchers remotely to rate each candidate topic on a 7-point scale across three dimensions: intensity of emotional arousal, threat to the relationships, and historical sensitivity. 

We averaged the scores from both partners across the three aspects to produce a final score for each topic. Crucially, topics with extreme scores were excluded; only topics with middle-range scores were selected to ensure the conflicts were manageable. If couples listed fewer than eight suitable topics, researchers provided a supplementary list of common relationship conflicts for them to rate and select from \cite{meyer2022relationship} (see Appendix \ref{apd:topics} for details). Before the simulated conflicts commenced, we rigorously balanced the conflict intensity across the assigned conditions to mitigate its confounding influence. As a result, the initial conflict intensities were highly balanced across all conditions ($M_{\text{baseline}} = 12.11$, $SD = 2.81$; $M_{\text{Basic Reminder}} = 12.36$, $SD = 3.93$; $M_{\text{Neutral Guide}} = 12.31$, $SD = 2.63$; $M_{\text{Empathetic Guide}} = 12.36$, $SD = 2.51$). A repeated measures ANOVA with a Greenhouse-Geisser correction confirmed that there was no significant difference in the initial conflict intensities among the conditions ($F(2, 34.07) = 0.08, p = .928$), indicating that the conflict difficulty was successfully controlled.

During each session, couples were instructed to recall specific points of disagreement regarding their assigned topic and discuss it via the developed web application, \textit{SpeakSoftly}. The discussion time was limited to 16 minutes, and the researcher gave a reminder when two minutes remained. If participants finished early, they were instructed to contact the researcher directly. Participants were asked to use devices with a 10-inch or larger monitor for all discussions during simulated conflict sessions to ensure full visibility of the AI suggestions alongside the chat interface. After each discussion, participants completed a post-session questionnaire. Following the final session, participants were interviewed.

\subsubsection{Real-Life Conflicts Phase}
The real-life conflict phase lasted for four weeks. Participants were asked to access the \textit{SpeakSoftly} web application on their preferred device whenever they subjectively perceived a conflict occurring in their daily lives. Usage was entirely autonomous and voluntary; participants were free to initiate discussions based on their own judgment, provided they were willing to share the data for research purposes, or revert to their accustomed approaches for any conversations they deemed too private to discuss via textual form. If participants autonomously engaged in three conflict discussions using the three intervention modes respectively during this period, their data were included in the analysis. To monitor psychological states and ensure accurate recall, participants were required to report to the researcher immediately and complete the post-session questionnaire on the same day the conflict occurred. After the third reporting, participants were interviewed.

\subsection{Evaluation}
\subsubsection{Evaluation Metrics}
\label{subsec:measures}
As applying NVC is an ongoing process aimed at fostering mutual understanding rather than merely forcing immediate conflict resolution, our evaluation focused on the multidimensional outcomes of the intervention. To rigorous measure these effects, we drew upon established psychological and HCI scales. All items were administered via a post-session questionnaire using a 7-point Likert scale (see Appendix \ref{apd:questionnaire} for the full list of items). Our measures were categorized into one system evaluation metric and three primary constructs: 
(1) \textbf{system acceptance and perceived effectiveness} (2 items); 
(2) \textbf{behavioral and interaction changes} (7 items), which encompasses both the reduction in verbal aggression and the enhancement of communication competence (adapted from the CCSS scale \cite{jones2018development}); 
(3) \textbf{cognitive and emotional shifts} (6 items), which captures both self-reflection and insight (adapted from the SRIS scale \cite{grant2002self}) and empathy and perspective-taking (adapted from the IRIC \cite{peloquin2010measuring} and CCSS \cite{jones2018development} scales); 
(4) \textbf{perceived change in relational quality} (3 items, adapted from \citet{scissors2016bias}).

Moreover, each phase concluded with a brief exit interview to gather subjective feedback. Participants were individually asked questions such as \textit{"Please rank the three intervention modes by your perceived effectiveness and explain your reasoning,"} \textit{"What positive or negative impacts, if any, did you notice on yourself, your partner, and your relationship during the phases?"} \textit{"Have you tried any communication techniques to resolve conflicts in the past? If so, how would you compare them to the interventions in this study?"} and \textit{"Was there any difference between your experiences in the simulated and real-life conflict phases?"}

\subsubsection{Data Analysis}
We adopted a mixed-methods approach to integrate quantitative statistical results with qualitative interview feedback. 
% 内部一致性检验
To validate our quantitative measures, we assessed internal consistency via Cronbach's alpha. The results demonstrated excellent reliability for the overall scale ($\alpha = .95$) as well as strong reliability for the primary constructs: behavioral and interaction changes ($\alpha = .92$), cognitive and emotional shifts ($\alpha = .83$), and perceived change in relational quality ($\alpha = .91$). 
% 正态性检验
The Shapiro-Wilk test indicated significant deviations from normality for all four constructs in at least one condition (\textit{p}s $<$ .05). Thus, we used the non-parametric Friedman test for main effects, followed by Wilcoxon signed-rank tests with Bonferroni correction for post-hoc pairwise comparisons.

Due to the privacy-preserving attrition described in Section~\ref{subsec:participants}, the real-life phase yielded a smaller sample size and was statistically underpowered. Furthermore, real-life conflicts inherently introduced uncontrolled variables, with participants reporting noticeably higher conflict intensity during these field scenarios ($M = 13.16$) compared to the simulated ones ($M = 12.20$). To ensure methodological rigor, all quantitative findings and statistical significance tests reported in our results are derived exclusively from the well-controlled simulated conflict phase. The qualitative data gathered during the real-life phase provided complementary, ecologically valid insights. In our analysis, we triangulate qualitative feedback from both phases to address not only what changes occurred, but also when and how the interventions were most effective during real-life, high-stress conflicts.

\subsection{Ethical and Safety Considerations}
Designing an LLM-powered intervention for couples experiencing conflict needs to address a number of ethical and safety concerns. We implemented a robust risk management plan tailored to the distinct challenges of our study. 

First, unconstrained text-based conflicts might be dangerous for partners' emotional well-being, as misunderstandings can easily escalate into verbal aggression. It is essential to ensure that the conflict topics and emotional intensity are manageable, which can be difficult if couples are exposed to unguided, extreme emotional triggers. To avoid this issue, we implemented strict preventive mechanisms during the simulated phase. We ensured that conflicts were grounded in carefully calibrated, middle-range intensity topics (scored 12.11-12.36 out of 21), deliberately excluding those that posed extreme threats to the relationship. Additionally, we introduced a one-day emotional washout period between sessions to minimize the compounding emotional impact of prior conflicts. 

Second, enabling free-form discussions about intimate relationships in the wild can lead to additional privacy and safety concerns. User inputs might contain sensitive personal information, and natural conflicts are inherently unconstrained. To address these risks, participant engagement in the real-life phase was strictly voluntary. This allowed us to avoid potentially sensitive disclosures, as participants retained full autonomy to bypass the textual interface and revert to their accustomed communication channels for any conversations they deemed too private. Furthermore, the nature of \textit{SpeakSoftly} inherently served as a real-time emotional safeguard, reminding users to pause and reflect before escalation. Additionally, to mitigate the risk of the LLM generating inappropriate advice, we strictly constrained the AI's behavior by prompt engineering, as detailed in Section~\ref{sec:implementation}. Finally, we required immediate reporting after real-life conflict usages, allowing researchers to continuously monitor psychological states. Throughout the study, no participants reported experiencing psychological distress or required study termination.

\section{Results}

\begin{figure}
    \centering
    \includegraphics[width=\textwidth]{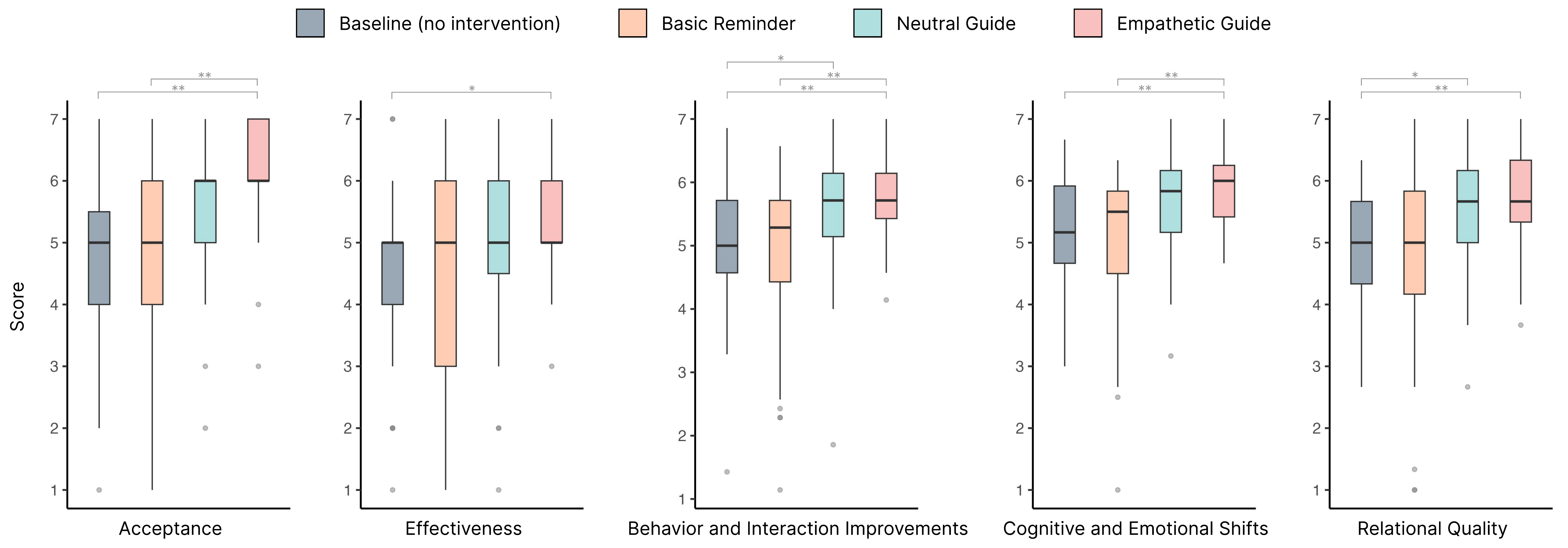}
    \caption{\textbf{Overview of Quantitative Results.} Box plots comparing scores for the Baseline (no intervention) (gray), Basic Reminder (orange), Neutral Guide (green), and Empathetic Guide (pink) across the five evaluation metrics. Asterisks denote statistical significance (* indicates $p < 0.05$, ** indicates $p < 0.01$, *** indicates $p < 0.001$).}
    \label{fig:quantitative}
    \Description{This image contains five separate box plots. Each plot compares the results of four different conditions, which are color-coded: "Baseline (no intervention)" (gray), "Basic Reminder" (orange), "Neutral Guide" (green), and "Empathetic Guide" (pink). The vertical axis for all plots is labeled "Score" and ranges from 1 to 7. The titles of the five plots are: System Acceptance, Perceived Effectiveness, Behavior and Interaction Changes, Cognitive and Emotional Shifts, and Perceived Relational Quality. Although median scores overlap or are identical in some categories, there is a clear upward trend in the overall rank distributions. The "Baseline (no intervention)" condition generally shows the lowest distributions. The ratings then increase for the "Basic Reminder" and "Neutral Guide" conditions, with the "Empathetic Guide" condition receiving the most favorable distributions across all categories. Brackets with asterisks denote statistically significant differences (* means p < 0.05, ** means p < 0.01, and *** means p < 0.001).}
\end{figure}

% 系统好不好用&接受度 → 表面行为改变了吗 → 深层认知改变了吗 → 最终关系变好了吗？
Building on the four primary evaluation constructs defined in Section~\ref{subsec:measures}, the following sections present our findings by integrating both quantitative and qualitative data. An overview of the quantitative results across all conditions is presented in Figure~\ref{fig:quantitative}. We structure our results to follow a progressive narrative: examining system acceptance and perceived effectiveness (Section~\ref{subsec:effectiveness}), analyzing surface-level communication shifts (Section~\ref{subsec:behavior}), exploring deeper psychological mechanisms (Section~\ref{subsec:cognitive}), and ultimately assessing the impact on the couples' relational quality (Section~\ref{subsec:relational}).
To understand not only \textit{what} changes occurred but also \textit{why} and \textit{how} the interventions facilitated these shifts, we triangulate the statistical results with interview feedback under each construct.

\subsection{System Acceptance and Effectiveness}
\label{subsec:effectiveness}

System acceptance scores across the four conditions were as follows: Empathetic Guide ($Mdn=6.00, IQR=1.00$), Neutral Guide ($Mdn=6.00, IQR=1.00$), Basic Reminder ($Mdn=5.00, IQR=2.00$), and baseline ($Mdn=5.00, IQR=1.50$). Although some median scores were identical, a Friedman Test revealed a significant difference in the rank distributions among the conditions, $\chi^2(3)=25.95, p<.0001$. Post-hoc Wilcoxon Signed-Rank Tests with Bonferroni correction showed that participants consistently rated the Empathetic Guide as significantly more acceptable than both the baseline ($p<.001$) and the Basic Reminder ($p=.002$). No other differences were statistically significant. 

A similar pattern was observed for perceived effectiveness: Empathetic Guide ($Mdn=5.00, IQR=1.00$), Neutral Guide ($Mdn=5.00, IQR=1.50$), Basic Reminder ($Mdn=5.00, IQR=3$) and baseline ($Mdn=5.00, IQR=1$). Although the median scores were identical, a Friedman Test revealed a significant difference in the rank distributions among the conditions, $\chi^2(3)=14.40, p<.001$. Post-hoc Wilcoxon Signed-Rank Tests with Bonferroni correction showed that the Empathetic Guide was rated significantly more effective than the baseline ($p=.017$). 

% Ranking of Effectiveness: Basic Reminder 4, Neutral Guide 8, Empathetic Guide 27

% why Empathetic is preferred
Aligned with the quantitative results, the majority of participants (70\%) voted the Empathetic Guide as the most effective mode. Participants frequently attributed this to its friendly, anthropomorphic tone and use of emojis, which made the behavioral suggestions feel warmer and easier to accept (P2-P4, P6-P8, P11, P13, P14, P17-P19, P21, P23-P26, P29-P32, P34). For instance, P18 noted: \textit{"The [Empathetic] approach is closer to real-life communication, making it easier to accept. It feels more relatable and has a certain degree of friendliness."} Furthermore, positive reinforcement mechanisms, especially the thumbs-up emojis feedback, provided a sense of encouragement (P8, P12, P13, P16, P19, P23, P26, P31, P32). P32 commented: \textit{"The thumbs-up feature encourages me to proactively avoid verbal aggression."} P12 similarly highlighted this motivational mechanism: \textit{"It feels almost like a reward system. It triggers a bit of competitiveness in me. I want to collect more thumbs-ups, which actively motivates me to do the right thing."}

% when Neutral is better
Conversely, a smaller subset of participants preferred the Neutral Guide (21\%) or the Basic Reminder (9\%). This preference emerged as particularly significant in emotionally heightened situations where simplicity was paramount. Several participants noted that during intense real-world conflicts, the longer text of the Empathetic Guide could be easily ignored (P9, P10, P17, P18, P19). As P18 articulated, while the Empathetic Guide was richer and more persuasive, the Neutral Guide excelled in real-life scenarios due to its brevity: \textit{"While the [Empathetic] guide has richer content that convinces me to rephrase my messages, the [Neutral] guide is slightly superior in terms of convenience and efficiency...When emotions run high during arguments, I may not have the patience to read long texts. The [Neutral] approach's brevity makes it easier to glance through and absorb the advice."} Finally, a few participants preferred the Basic Reminder. P1 explained that overly detailed interventions might cause users to \textit{"lose our own words, thoughts, and personality in communication."} P10 added that having too many guided features becomes ineffective when users are experiencing intense negative emotions.

\subsection{Behavioral and Interaction Changes}
\label{subsec:behavior}
\begin{figure}
    \centering
    \includegraphics[width=\textwidth]{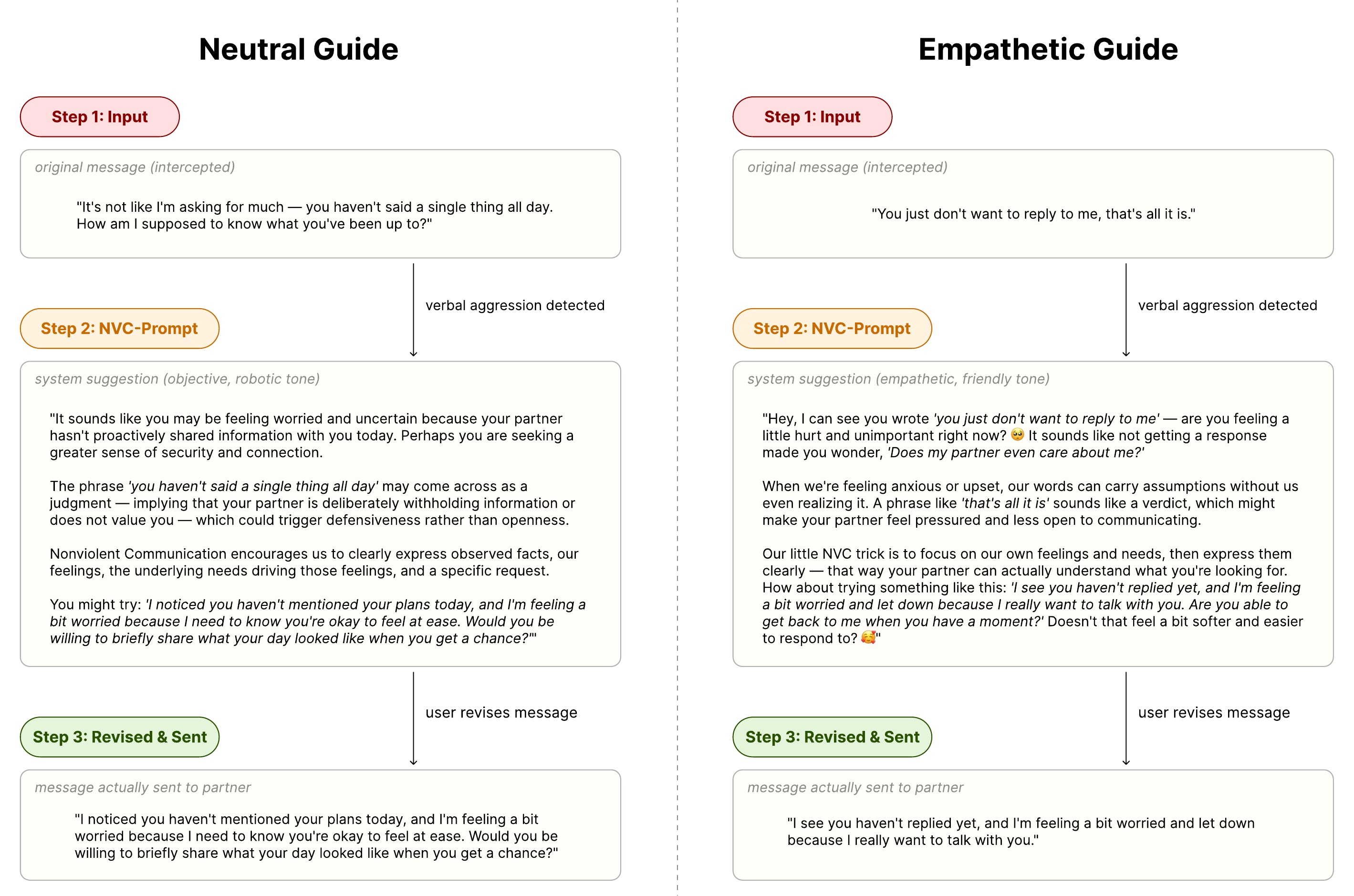}
    \caption{\textbf{Real Interaction Examples of NVC-Prompt Interception.} Two representative cases illustrating how the NVC-Prompt operated under the Neutral Guide (left) and Empathetic Guide (right) during simulated conflicts. Each case shows the user's original message that triggered interception, the system's tailored suggestion, and the user's revised message. Messages are paraphrased from actual participant interactions to preserve anonymity.}
    \label{fig:realexample}
    \Description{Two-panel figure showing the NVC-Prompt interception process. The left panel demonstrates the Neutral Guide mode: a user's original message expressing frustration about their partner not sharing daily plans is intercepted, the system provides a structured suggestion analyzing the user's feelings, identifying judgmental language, summarizing NVC principles, and offering a revised phrasing, and the user sends a revised message. The right panel demonstrates the Empathetic Guide mode: a user's original message accusing their partner of not wanting to reply is intercepted, the system provides a warm and empathetic suggestion that validates the user's emotions, gently identifies assumptions in the language, and offers a softer alternative phrasing, and the user sends a revised message. Both panels follow a three-step vertical flow connected by downward arrows labeled with the triggering action.}
\end{figure}

Behavior and interaction improvement scores across the four conditions were as follows: Empathetic Guide ($Mdn=5.71, IQR=0.71$), Neutral Guide ($Mdn=5.71, IQR=1.00$), Basic Reminder ($Mdn=5.29, IQR=1.29$), and baseline ($Mdn=5.00, IQR=1.14$). Although the highest median scores were identical, a Friedman Test revealed a significant difference in the rank distributions among the conditions, $\chi^2(3)=20.94, p<.001$. Post-hoc Wilcoxon Signed-Rank Tests with Bonferroni correction showed that participants reported significantly greater behavioral and interaction improvements with the Empathetic Guide compared to both the baseline ($p=.002$) and the Basic Reminder ($p=.003$). Similarly, the Neutral Guide resulted in significantly greater improvements than the baseline ($p=.012$). No other differences were statistically significant. These results indicate that providing concrete rephrasing guidance, regardless of whether the tone is empathetic or neutral, significantly reduces surface-level verbal aggression and enhances communication competence. In contrast, merely alerting users (Basic Reminder) was insufficient to trigger significant behavioral changes. To illustrate this process in practice, Figure~\ref{fig:realexample} presents two representative interception cases from the Neutral Guide and Empathetic Guide, showing how users revised their messages after receiving NVC-Prompt suggestions. Qualitative feedback further illuminated the mechanisms behind these behavioral changes.

\textbf{Conversational friction introduced space for emotional regulation.}
Qualitative feedback illuminated how this concrete guidance facilitated behavioral changes. A primary mechanism was the introduction of a "just-in-time" pause that forced emotional regulation. Participants reported that the NVC-Prompt created a physical and temporal friction preventing the immediate sending of impulsive messages which provided the necessary space to calm down (P3, P5, P6, P9, P11-P14, P18, P20, P22-P26, P30, P31-P33, P36). As P31 described, \textit{"When I was angry and thought my partner was being unreasonable, I wanted to send a very aggressive message. But the pop-up interrupted me and advised against using violence on my partner, which successfully stopped my impulse."} Rather than using the chat as a cathartic outlet, this induced pause successfully mitigated the competitive mindset typical of conflicts, fostering a shift toward rational communication (P18, P30). The Empathetic Guide was particularly effective in this regard; P14 explained: \textit{"When emotions run high during conflicts, it is so easy to type out angry words. The system forced me to look at a warm, vivid reminder. After reading it, my emotions cooled down, preventing a verbal outburst."} 

\textbf{Real-time feedback revealed blind spots in self-expression.}
Beyond merely delaying the message, the guidance helped users recognize the unintended severity of their own wording. Participants noted that in standard messaging apps, they often focus solely on their own frustration, oblivious to how their partner might interpret their words (P2, P14, P24). P2 reflected: \textit{"Usually, when arguing on WeChat without any reminders, you only focus on yourself and don't realize if the sentence you just typed will hurt the other person."} The NVC-Prompt successfully bridged this blind spot. P14 shared a realization: \textit{"I didn't think my words were too harsh, but the assistant pointed out they could be very heavy for my partner. I remembered this, and consciously avoided those hurtful phrases in subsequent conflicts."}

\textbf{Scaffolded guidance transformed paused thoughts into constructive expression.}
Finally, participants found both system components instrumental in transforming their paused thoughts into constructive expressions by acting as a linguistic scaffolding tool. Rather than blindly adopting the provided suggestions, users leveraged the generated text from both the NVC-Prompt and NVC-Guide to reconstruct their messages. 
For immediate phrasing corrections, the NVC-Prompt provided direct, usable alternatives. Several participants noted that the gentle rephrasings offered could be referenced to soften their tone on the spot (P8, P14, P16, P26). As P16 explained: \textit{"The prompt showed me how to rephrase things I would normally say harshly. I intentionally tried using some of the aggressive language I would use in real conflicts, and the system taught me alternative ways to express the same thing. That was the most valuable learning for me."}
For deeper dialogue reconstruction, the NVC-Guide offered contextual scaffolding. P36 exemplified: \textit{"The NVC-Guide points out details I might have missed based on my conversations. It helps me extract useful phrases to communicate better...While I don't use the suggestions verbatim, I incorporate them into my own responses, which has significantly reduced the likelihood of arguments"}. Similarly, P11 noted that the NVC-Guide helped them better articulate their thoughts while ensuring the phrasing was highly acceptable and less likely to be misinterpreted by their partner.

\subsection{Cognitive and Emotional Shifts}
\label{subsec:cognitive}
Cognitive and emotional shift scores across the four conditions were as follows: Empathetic Guide ($Mdn=6.00, IQR=0.83$), Neutral Guide ($Mdn=5.83, IQR=1.00$), Basic Reminder ($Mdn=5.50, IQR=1.33$), and baseline ($Mdn=5.17, IQR=1.25$). A Friedman Test revealed a significant difference among the conditions, $\chi^2(3)=16.04, p=.001$. Post-hoc Wilcoxon Signed-Rank Tests with Bonferroni correction showed that participants reported significantly deeper cognitive and emotional shifts with the Empathetic Guide compared to both the baseline ($p=.008$) and the Basic Reminder ($p=.003$). No other differences were statistically significant. 

These quantitative results highlight a critical divergence: while the Neutral Guide significantly improves behavioral and interaction patterns (as shown in Section~\ref{subsec:behavior}), it is often insufficient to trigger internal mindset changes. Qualitative findings reveal that fostering deep cognitive shifts fundamentally requires an empathetic approach. The warm, anthropomorphic tone of the Empathetic Guide uniquely dismantles users' psychological defenses, enabling them to connect emotionally with their partner's feelings. As P34 shared: \textit{"The [Neutral Guide] approach feels rigid, like solving a problem step-by-step without really touching my heart. While I understand what to do, it feels less personal and engaging compared to the Empathetic Guide."} Similarly, P20 emphasized how this tone helped perspective-taking: \textit{"I’m not someone who naturally considers others’ perspectives. But the anthropomorphic tone of the intervention reminded me to think about what the other person's feeling."} Building on this emotional connection, the Empathetic Guide and NVC-Guide facilitated cognitive transformations in two directions: an inward reflection on one's own true feelings, and an outward perspective-taking of their partner's needs. 

\textbf{Emotional regulation prompted reflection on the root causes of aggression.}
By prompting emotional regulation, the system successfully guided users to reflect on the root causes of their aggressive behaviors rather than simply venting anger (P13, P18, P19, P22, P27, P28, P30). When individuals are caught in a heated conflict, their true needs are often masked by aggressive rhetoric. The NVC-Guide effectively pierced through this anger to reveal users' underlying vulnerabilities. P13 described this sudden realization: \textit{"The [NVC-Guide] function is excellent. It analyzes the true intentions behind the harsh words I just typed. The intentions I wasn't even aware of myself. Once it points them out, I have a sudden realization. My anger instantly dissipates, and I start to genuinely reflect on and address the root cause of my aggression."} Similarly, P28 noted that the NVC-Guide's analysis of their own emotional state forced them to pause and recalibrate, moving from a state of blind output to mindful self-awareness.

\textbf{Shifting from adversarial arguing to mutual understanding.}
Beyond self-reflection, the interventions significantly enhanced users' ability to step into their partner’s shoes (P3-P6, P9, P13, P16, P17, P20-P23, P27, P28, P30, P32, P35). Participants noted that conflicts typically trigger a competitive mindset where both parties only want to be heard, not to listen. P5 captured how the system disrupted this pattern: \textit{"When you're arguing, you usually just want the other person to understand you, but you don't think about understanding them. But the intervention reminds you to pause and think, 'Am I being too harsh on my partner? Should I reflect on my own behavior?'"} Furthermore, the contextual analysis of NVC-Guide helped participants uncover the hidden emotional labor and unmet needs of their partners, thereby de-escalating the conflict at a relational level (P4, P5, P13, P16, P32, P36). P27 shared a realization: \textit{"The analysis made me much more empathetic. I suddenly realized that she is also trying so hard to prevent us from fighting, and she is making an effort to understand me. I hadn't thought that deeply before...I realized I was often just venting my negative emotions on her. Having the assistant intervene and remind me of this was a wake-up call."} By surfacing these previously unspoken needs (as P17 also noted, bringing "hidden topics to the table"), the system effectively transformed adversarial arguments into opportunities for mutual empathy.

\subsection{Impact on Relational Quality and Real-World Applicability}
\label{subsec:relational}
Perceived relational quality scores across the four conditions were as follows: Empathetic Guide ($Mdn=5.67, IQR=1.00$), Neutral Guide ($Mdn=5.67, IQR=1.17$), Basic Reminder ($Mdn=5.00, IQR=1.67$), and baseline ($Mdn=5.00, IQR=1.33$). A Friedman Test revealed a significant difference among the conditions, $\chi^2(3)=18.30, p<.001$. Post-hoc Wilcoxon Signed-Rank Tests with Bonferroni correction showed that both the Empathetic Guide ($p=.004$) and the Neutral Guide ($p=.047$) were rated significantly higher in perceived change in relational quality than the baseline. No other differences were statistically significant. 

These quantitative results suggest that both the explicit behavioral improvements (facilitated by the Neutral Guide or Empathetic Guide) and the implicit cognitive shifts (fostered by the Empathetic Guide) successfully translated into enhanced relational quality. Qualitative feedback aligned with this pattern, indicating that the interventions actively repaired and strengthened intimate bonds rather than merely preventing arguments. P6 explicitly noted that their \textit{"intimate relationship became stronger"} as a result of the intervention. P20 further elaborated on how emotional regulation and perspective-taking fostered long-term relational optimism: \textit{"First, I become less agitated, and then I find myself stepping into my partner's shoes to understand his point of view. Our relationship will progressively get better because our communication has become much more effective and constructive than before."}

Beyond improved relational quality within the study, qualitative feedback suggested that the interventions may have encouraged more mindful communication habits in participants' daily lives (P6, P11, P12, P18, P19, P26, P36). As P12 shared: \textit{"The [NVC-Prompt] reminders made me more mindful in daily life. I became more conscious of my word choices and realized that some things just shouldn't be said."} Furthermore, participants contrasted this active habit-building with passive learning from social media (P14, P19, P20). P20 contrasted the two approaches: \textit{"With tips from social media, I can keep it up for a couple of days, but then I revert to old habits. But the AI [intervention] is different, it keeps reminding me and giving me concrete suggestions whenever I slip...Social media relies entirely on willpower, which usually lasts a day or two at most."} P34 was even motivated to independently research NVC principles.

Participants also noted that the JIT mechanism addressed critical shortcomings of general-purpose conversational agents such as Doubao\footnote{Doubao is a conversational AI assistant developed by ByteDance, similar to ChatGPT.} (P8, P11, P12, P23, P25, P26). A key issue with post-hoc AI analysis is the distortion of context: consulting a chatbot after an argument requires users to rephrase or summarize the event, which inherently loses the raw emotional intensity and objectivity of the original situation (P8, P11, P12). P32 further emphasized the granularity of intervention: \textit{"While post-hoc chatbots provide a holistic summary of the conflict, our platform offers real-time, sentence-by-sentence guidance, preventing the damage before it occurs."} Beyond these technical advantages, P34 valued the system's role as an objective mediator within the couple's private boundary, offering impartial guidance without exposing their conflicts to friends or family. 
\section{Discussion}

\subsection{Empathetic Tone as the Catalyst for Deep Cognitive Shift}

Based on our quantitative results and qualitative findings from simulated conflicts, both Empathetic Guide and Neutral Guide, which share the same NVC scaffolding content, significantly reduced verbal aggression. However, only Empathetic Guide produced significant cognitive and emotional shifts, including enhanced self-awareness and perspective-taking. This finding suggests that structured scaffolding content is a necessary but not sufficient condition for cognitive shift. We interpret this through the lens of self-determination theory \cite{ryan2024self}: empathetic tone fosters a sense of relatedness (the fundamental need to feel connected to and understood by others), which in turn promotes intrinsic motivation that supports sustained engagement and internalization. As \citet{vansteenkiste2020basic} argue, when guidance is delivered through a warm and supportive character that conveys genuine connection, the goals and behaviors being encouraged are more likely to be deeply internalized rather than merely complied with. This explains why Neutral Guide, despite providing the same instructional content, changed what users said but not how they felt: its emotionally distant tone failed to activate the relatedness needed for genuine cognitive restructuring.

Our findings also provide empirical validation for the Needs-Conscious Design framework proposed by \citet{wolfe2025toward}. NVC-Prompt enacts Intentionality by slowing down fast-paced digital communication to foster self-reflection, while NVC-Guide enacts Receptiveness to Needs by shifting users' focus from assigning blame toward uncovering unmet needs. Beyond validating these pillars, our results reveal a critical dimension that the original framework does not address: the role of tone in activating receptiveness. Without the sense of relatedness and psychological connection facilitated by empathetic tone, users may understand what they should do differently yet remain unwilling to internalize that change at the cognitive and emotional level. This is because the intrinsic motivation to do so has not been activated, we therefore propose that empathetic tone be recognized as a necessary complementary dimension of the Needs-Conscious Design framework.

\subsection{Design Implications for In-Situ Interventions in High-Stakes Communication}

\subsubsection{Friction as a Feature, Not a Bug}

A common principle in communication technology design is to minimize friction, enabling messages to be sent as quickly and seamlessly as possible. Yet in the context of intimate conflict, this seamlessness may be precisely the problem: it allows impulsive, emotionally charged messages to reach a partner before the sender has had a chance to reconsider. As \citet{roloff1987interpersonal} observed, escalating conflicts often lead to improvisational responses that individuals later regret. Our JIT intervention directly counters this dynamic. By intercepting a message flagged for verbal aggression and presenting a pop-up before it is delivered, NVC-Prompt introduces a deliberate pause into the communication flow. Rather than being an obstacle, this friction functions as a protective mechanism, aligning what \citet{cox2016friction} characterize as a productive disruption that transforms mindless interaction into a moment of reflection.

Our three intervention modes reveal that the effectiveness of such friction depends not on its mere presence, but on its depth. Basic Reminder introduces the lightest frictionyet produced no significant improvement over baseline. This suggests that simply interrupting the flow is insufficient; users need more than a signal to stop, they need guidance on what to do next. Neutral Guide and Empathetic Guide, which embed structured NVC-based revision suggestions within the pause, both produced significant behavioral changes. Empathetic Guide went further, uniquely producing significant cognitive and emotional shifts as well. Through the lens of Gross's process model of emotion regulation \cite{gross2002emotion}, we interpret this progression as follows: the pop-up creates a temporal window for reappraisal, an early emotion regulation strategy that works by changing how a situation is construed before the emotional response fully forms. But this window is only useful if filled with substantive content that actively guides the reappraisal process. Empathetic Guide succeeds because it does not merely instruct users to revise their words; it models an empathetic perspective that reshapes how users interpret the conflict itself. We elaborate on the calibration of friction depth to emotional intensity in Section~\ref{subsec:implications:adaptive}.

This reappraisal-oriented design stands in deliberate contrast to suppression, the other major down-regulation strategy in Gross's framework \cite{gross2002emotion}. Suppression occurs later in the emotion-generative process: rather than changing how a situation is interpreted, it requires individuals to inhibit the outward expression of already-formed emotions. Our formative interviews illustrated this distinction. Participants described the immense self-regulation required when attempting to apply communication techniques independently, frequently reporting a feeling of "swallowing grievances" and suppressing what they truly felt in order to appear calm. \citet{gross2002emotion} demonstrates that while suppression may reduce behavioral expression, it fails to decrease the underlying emotional experience and actually impairs memory. By intervening before a destructive message is sent and providing structured guidance within that pause simultaneously, NVC-Prompt steers users toward reappraisal over suppression. Without such scaffolding, the pause alone risks becoming another occasion for suppression, where users refrain from sending but remain emotionally unchanged. This early intervention coupled with substantive guidance, not only produces more genuine behavioral change but also lays the groundwork for the kind of sustained cognitive shift and habit formation we discuss in the following section.

\subsubsection{Scaffolding over Automation}

A truly effective intervention should scaffold users toward growth rather than automate communication on their behalf. Although our system provides revision suggestions, it deliberately avoids offering a one-click replacement button, instead acting as what \citet{fu2024text} describe as a "wise friend," guiding users while preserving their autonomy. Crucially, users retain full control: a "skip" button allows them to dismiss the pop-up and send their original message unchanged, ensuring that the intervention remains a suggestion rather than a mandate. Our qualitative findings confirm that participants engaged actively with this design: rather than copying and pasting suggestions, they reconstructed their intercepted messages in their own words based on the suggested direction (see Figure~\ref{fig:realexample}). This reconstruction process requires users to reflect on why their original message was flagged and to develop their own understanding of the underlying communication principles, ensuring that the effort remains authentically the sender's own. In this way, our design avoids the risk of "empathy fog" (e.g., hollow or inauthentic messages) \cite{wolfe2025toward}, because the final message is genuinely crafted by the user.

Moreover, this scaffolding mechanism extends beyond the immediate conflict toward the establishment of long-term communication habits. Some participants reported that the interventions carried over into daily life, making them more mindful of their word choices even outside the platform as they came to recognize how words can cause harm. This is particularly notable because most digital behavior change interventions overlook habitual behavior \cite{pinder2018digital}, yet destructive communication patterns (which often trigger emotional reactions \cite{gottman2023predicts}) tend to be enacted without purposeful thinking or awareness \cite{nilsen2012creatures}, and are therefore difficult to change. Our intervention addresses this by functioning as a consistent external trigger that disrupts the automatic cycle. Drawing on \citet{duhigg2012power}'s habit loop framework, which emphasizes the role of triggers, behaviors, and rewards in habit formation, we interpret NVC-Prompt as the trigger, the reconstruction behavior as the new routine, and both the explicit positive reinforcement feedback in Empathetic Guide (e.g., thumbs-up acknowledgment) and the resulting improvement in conversational quality as the reward. Through this repeated cycle, users gradually internalize healthier mindsets, fostering self-awareness and perspective-taking.

\subsubsection{AI as a Private-Sphere Mediator}

Previous work on mediating intimate conflicts has explored chatbots as explicit third parties that communicate directly with individuals \cite{zheng2021pocketbot, yuksel2022conversational}. Our design takes a fundamentally different approach: rather than acting as a visible mediator who engages both partners in dialogue, \textit{SpeakSoftly} operates as an embedded, private assistant that each individual interacts with independently within the conversation flow. This distinction matters because introducing a third party into intimate conflicts carries inherent risks. As \citet{rubin1985third} observed, third parties in family conflicts can disrupt dyadic stability and trigger competitive dynamics, particularly when one partner perceives the third party as taking sides. Our participants echoed this concern, P33 expressed a desire to share chat history with a friend to determine who was "right," but her partner opposed this due to privacy concerns. Participants valued that our intervention operated entirely within the private sphere, with no external audience and no risk of judgment being disclosed. In this way, the AI functions as what \citet{rubin1985third} described as a face-saving mechanism: users can soften their language and reconsider their stance without this shift being attributable to weakness, since it is framed as responding to a system suggestion rather than capitulating to the partner.

\subsubsection{Adaptive Intervention for Emotional Intensity}
\label{subsec:implications:adaptive}

Interestingly, although Empathetic Guide was the most effective mode in simulated conflicts, participants reported that Neutral Guide was more practical in real-life settings. They explained that the empathetic, emotionally rich language in Empathetic Guide required more cognitive effort to process than the direct, matter-of-fact tone of Neutral Guide. This is not because the suggestions were substantially longer, but because the emotional framing itself demanded additional processing. This pattern aligns with cognitive load theory: participants rated emotional intensity higher in real-life conflicts than in simulated ones, and as \citet{plass2019four} argue, such heightened emotions impose extraneous cognitive load by competing for the limited capacity of working memory. Under these conditions, the richer suggestions of Empathetic Guide likely exceeded users' available processing resources, whereas Neutral Guide's directness better matched their constrained cognitive capacity.

These findings suggest that no single intervention depth suits all contexts, particularly in high-stakes communication where emotional intensity fluctuates throughout a conversation. An effective intervention should adapt its depth to the user's current emotional state. For example, a future system could estimate emotional intensity from signals such as the frequency and severity of verbal aggression in recent messages, delivering Empathetic Guide's warm, friend-like scaffolding during lower-intensity moments while switching to Neutral Guide's concise, professional tone when emotions peak. Such adaptive calibration would ensure that the intervention remains within users' cognitive capacity at all times, maximizing both effectiveness and acceptance.

\subsection{Designing JIT Interventions for Broader Communication Contexts}

While our empirical findings are situated within the context of intimate couple conflicts, NVC itself is an approach designed for all levels of communication and diverse situations, including families, schools, organizations, and diplomatic negotiations \cite{rosenberg2015nonviolent}. We believe the underlying rationale of our system design may hold relevance beyond this specific context. At the framework level, parent–child conflicts during homework scenarios present similar dynamics of escalation and emotional reactivity \cite{gao2025homework}, and workplace communication could likewise benefit from real-time nudges toward constructive expression. It could also support individuals with tendencies toward verbal aggression more broadly, helping them practice self-regulation and develop healthier communication habits. Beyond expanding to new populations, our design could also be adapted to richer communication modalities \cite{scissors2013back}. For example, pop-up reminders appearing on a user's screen during remote meetings to offer real-time feedback on their language. While such prompts cannot retract what has already been said, they may still meaningfully shape subsequent behavior. These extensions, however, remain speculative and require dedicated empirical validation in each new context before any generalizability claims can be made.

\subsection{Limitations and Future Work}
This study has several limitations. 
First, the population primarily consisted of young couples who self-reported their conflict frequency and expressed a willingness to improve their relationships. Although this group represents the primary target users of our interventions, findings may not generalize to other age groups, relationship types, or individuals who are less motivated to change. Future work should examine interventions across more diverse populations.

Second, although participants were encouraged to use the interventions during real-life conflicts, the opt-in nature of this phase resulted in a smaller sample size, limiting the statistical power of quantitative comparisons in naturalistic settings. Qualitative findings from real-life usage were nonetheless incorporated throughout our analysis. Additionally, while the intervention demonstrates short-term carryover effects, longitudinal studies spanning several months or more are needed to assess whether the observed behavioral changes persist and deepen over time. 

Third, our system operates as a web application, requiring couples to migrate their conversations away from their accustomed communication platforms. Several participants noted that this platform-switching friction poses a significant barrier to real-world adoption, as individuals in heated arguments are unlikely to redirect their conversation to a separate tool (P33, P34, P36). Additionally, despite the absence of culturally embedded expressive modalities such as custom stickers (P9, P16, P18, P27, P28), the intervention still produced significant effects, suggesting that the core NVC scaffolding mechanism is robust even under suboptimal platform conditions. Future work should explore integrating such interventions directly into existing messaging platforms, which would simultaneously reduce the adoption barrier and enable richer multimodal expression.

Finally, the LLM-generated suggestions exhibited repetitiveness in prolonged interactions, with similar phrasings recurring across different stages of a single conversation (P5, P27). This study was conducted between June and September 2025, when the underlying language models were less capable than current state-of-the-art models. Advances in LLM reasoning, contextual awareness, and alignment may substantially mitigate this limitation.

\section{Conclusion}

To scaffold the application of communication techniques during conflict, we present \textit{SpeakSoftly}, a system grounded in Nonviolent Communication (NVC) that provides just-in-time interventions to intercept verbal aggression and guide users through NVC principles step by step toward establishing healthier communication habits. To evaluate the appropriate level of intervention depth, three progressive intervention modes: Basic Reminder, Neutral Guide, and Empathetic Guide. These modes were compared in a mixed-methods user study across simulated and real-life couple conflicts. Our results showed that Empathetic Guide was the most accepted and effective mode, significantly reducing users' verbal aggression while enhancing self-awareness and perspective-taking in simulated conflicts. In real-life conflicts, however, Neutral Guide was reported to be more practical due to the lower cognitive load it demands in emotionally heightened scenarios, highlighting the importance of adapting intervention depth to emotional intensity in high-stakes contexts. We further discuss why empathetic tone is a necessary condition for cognitive shift and propose design implications for in-situ interventions in high-stakes communication contexts. While our study focused on couples, the underlying design rationale may generalize to other intimate relationships and high-stakes communication contexts. Future work should explore adaptive, context-aware interventions and assess long-term effectiveness across diverse populations.

\section*{Acknowledgments of AI Use}
We used AI, specifically LLMs, during the system implementation, as detailed in Section~\ref{sec:implementation}. Also, we used LLMs to assist with grammar checking, language polishing, and figure generation. The authors take responsibility for the accuracy and integrity of all content in this paper.

\bibliographystyle{ACM-Reference-Format}
\bibliography{ref}

@article{meyer2022relationship,
  title={The relationship between conflict topics and romantic relationship dynamics},
  author={Meyer, Dixie and Sledge, Renata},
  journal={Journal of Family Issues},
  volume={43},
  number={2},
  pages={306--323},
  year={2022},
  publisher={Sage Publications Sage CA: Los Angeles, CA}
}

@inproceedings{scissors2016bias,
  title={On the bias: Self-esteem biases across communication channels during romantic couple conflict},
  author={Scissors, Lauren and Gergle, Darren},
  booktitle={Proceedings of the 19th ACM Conference on Computer-Supported Cooperative Work \& Social Computing},
  pages={383--393},
  year={2016}
}

@article{afifi2012standards,
  title={The “standards for openness hypothesis” why women find (conflict) avoidance more dissatisfying than men},
  author={Afifi, Tamara D and Joseph, Andrea and Aldeis, Desiree},
  journal={Journal of Social and Personal Relationships},
  volume={29},
  number={1},
  pages={102--125},
  year={2012},
  publisher={Sage Publications Sage UK: London, England}
}

@article{jones2018development,
  title={Development and validation of the couple communication satisfaction scale},
  author={Jones, Adam C and Jones, Rebecca Lucero and Morris, Neli},
  journal={The American Journal of Family Therapy},
  volume={46},
  number={5},
  pages={505--524},
  year={2018},
  publisher={Taylor \& Francis}
}

@article{grant2002self,
  title={The self-reflection and insight scale: A new measure of private self-consciousness},
  author={Grant, Anthony M and Franklin, John and Langford, Peter},
  journal={Social Behavior and Personality: an international journal},
  volume={30},
  number={8},
  pages={821--835},
  year={2002},
  publisher={Scientific Journal Publishers}
}

@article{peloquin2010measuring,
  title={Measuring empathy in couples: Validity and reliability of the interpersonal reactivity index for couples},
  author={P{\'e}loquin, Katherine and Lafontaine, Marie-France},
  journal={Journal of personality assessment},
  volume={92},
  number={2},
  pages={146--157},
  year={2010},
  publisher={Taylor \& Francis}
}

@inproceedings{ortloff2025small,
  title={Small, Medium, Large? A Meta-Study of Effect Sizes at CHI to Aid Interpretation of Effect Sizes and Power Calculation},
  author={Ortloff, Anna-Marie and Martius, Florin and Meier, Mischa and Raimbault, Theo and Geierhaas, Lisa and Smith, Matthew},
  booktitle={Proceedings of the 2025 CHI Conference on Human Factors in Computing Systems},
  pages={1--28},
  year={2025}
}

@article{skinner1963operant,
  title={Operant behavior.},
  author={Skinner, Burrhus F},
  journal={American psychologist},
  volume={18},
  number={8},
  pages={503},
  year={1963},
  publisher={American Psychological Association}
}

@book{gottman2007marriages,
  title={Why marriages succeed or fail},
  author={Gottman, John Mordechai and Gottman, John},
  year={2007},
  publisher={A\&C Black}
}

@article{sowan2024conflict,
  title={A conflict escalation comparison: Couples from the general population and couples engaged in high-intensity conflict},
  author={Sowan, Wafaa},
  journal={Family Relations},
  volume={73},
  number={2},
  pages={858--873},
  year={2024},
  publisher={Wiley Online Library}
}

@article{eaker2007marital,
  title={Marital status, marital strain, and risk of coronary heart disease or total mortality: the Framingham Offspring Study},
  author={Eaker, Elaine D and Sullivan, Lisa M and Kelly-Hayes, Margaret and D’Agostino Sr, Ralph B and Benjamin, Emelia J},
  journal={Psychosomatic medicine},
  volume={69},
  number={6},
  pages={509--513},
  year={2007},
  publisher={LWW}
}

@article{johnston1994high,
  title={High-conflict divorce},
  author={Johnston, Janet R},
  journal={The future of children},
  pages={165--182},
  year={1994},
  publisher={JSTOR}
}

@article{ames1982two,
  title={When two wrongs make a right: Promoting cognitive change by social conflict.},
  author={Ames, Gail J and Murray, Frank B},
  journal={Developmental Psychology},
  volume={18},
  number={6},
  pages={894},
  year={1982},
  publisher={American Psychological Association}
}

@article{markman1993preventing,
  title={Preventing marital distress through communication and conflict management training: A 4-and 5-year follow-up.},
  author={Markman, Howard J and Renick, Mari Jo and Floyd, Frank J and Stanley, Scott M and Clements, Mari},
  journal={Journal of consulting and clinical psychology},
  volume={61},
  number={1},
  pages={70},
  year={1993},
  publisher={American Psychological Association}
}

@book{winstok2012partner,
  title={Partner violence: A new paradigm for understanding conflict escalation},
  author={Winstok, Zeev},
  year={2012},
  publisher={Springer science \& Business media}
}

@book{hendrick2000close,
  title={Close relationships: A sourcebook},
  author={Hendrick, Clyde and Hendrick, Susan S},
  year={2000},
  publisher={Sage publications}
}

@article{johnson2016computer,
  title={How do computer-mediated channels negatively impact existing interpersonal relationships},
  author={Johnson, AJ and Bostwick, E and Anderson, C and Gilchrist-Petty, E and Long, S},
  journal={Contexts of the dark side of communication},
  pages={241--252},
  year={2016},
  publisher={Peter Lang New York, NY}
}

@article{kelly2018perceived,
  title={Perceived miscommunication in friends’ and romantic partners’ texted conversations},
  author={Kelly, Lynne and Miller-Ott, Aimee E},
  journal={Southern Communication Journal},
  volume={83},
  number={4},
  pages={267--280},
  year={2018},
  publisher={Taylor \& Francis}
}

@article{williamson2016effects,
  title={Effects of relationship education on couple communication and satisfaction: A randomized controlled trial with low-income couples.},
  author={Williamson, Hannah C and Altman, Noemi and Hsueh, JoAnn and Bradbury, Thomas N},
  journal={Journal of consulting and clinical psychology},
  volume={84},
  number={2},
  pages={156},
  year={2016},
  publisher={American Psychological Association}
}

@book{rosenberg2015nonviolent,
  title={Nonviolent communication: A language of life: Life-changing tools for healthy relationships},
  author={Rosenberg, Marshall B and Chopra, Deepak},
  year={2015},
  publisher={PuddleDancer Press}
}

@inproceedings{li2018review,
  title={Review of unconventional user interfaces for emotional communication between long-distance partners},
  author={Li, Hong and H{\"a}kkil{\"a}, Jonna and V{\"a}{\"a}n{\"a}nen, Kaisa},
  booktitle={Proceedings of the 20th International Conference on Human-Computer Interaction with Mobile Devices and Services},
  pages={1--10},
  year={2018}
}

@inproceedings{yang2017communicating,
  title={Communicating through a telepresence robot: A study of long distance relationships},
  author={Yang, Lillian and Neustaedter, Carman and Schiphorst, Thecla},
  booktitle={Proceedings of the 2017 CHI conference extended abstracts on human factors in computing systems},
  pages={3027--3033},
  year={2017}
}

@article{zheng2021pocketbot,
  title={" PocketBot Is Like a Knock-On-the-Door!": Designing a Chatbot to Support Long-Distance Relationships},
  author={Zheng, Qingxiao and Markazi, Daniela M and Tang, Yiliu and Huang, Yun},
  journal={Proceedings of the ACM on Human-Computer Interaction},
  volume={5},
  number={CSCW2},
  pages={1--28},
  year={2021},
  publisher={ACM New York, NY, USA}
}

@inproceedings{VR-LDR,
  title={TogetherReflect: Supporting Emotional Expression in Couples Through a Collaborative Virtual Reality Experience},
  author={Wagener, Nadine and Albensoeder, Daniel Christian and Reicherts, Leon and Wo{\'z}niak, Pawe{\l} W and Rogers, Yvonne and Niess, Jasmin},
  booktitle={Proceedings of the 2025 CHI Conference on Human Factors in Computing Systems},
  pages={1--16},
  year={2025}
}

@article{adebayo2024role,
  title={Role of Third Party Interference in Marital Stability among Married adults in Nigeria},
  author={Adebayo, David Obafemi and Michael, Ifeoluwa Blessing and Abisoye, Faysol Aderibigbe},
  journal={Buletin Konseling Inovatif},
  volume={4},
  number={3},
  pages={214--222},
  year={2024}
}

@article{a2000social,
  title={The social context of couple conflict: Support and criticism from informal third parties},
  author={A. Klein, Renate C and Milardo, Robert M},
  journal={Journal of Social and Personal Relationships},
  volume={17},
  number={4-5},
  pages={618--637},
  year={2000},
  publisher={Sage Publications Sage CA: Thousand Oaks, CA}
}

@article{rubin1985third,
  title={Third party intervention in family conflict},
  author={Rubin, Jeffrey Z},
  journal={Negotiation Journal},
  volume={1},
  number={3},
  pages={269--281},
  year={1985},
  publisher={MIT Press 255 Main Street, 9th Floor, Cambridge, Massachusetts 02142, USA~…}
}

@inproceedings{scissors2013back,
  title={" Back and forth, back and forth" channel switching in romantic couple conflict},
  author={Scissors, Lauren E and Gergle, Darren},
  booktitle={Proceedings of the 2013 conference on Computer supported cooperative work},
  pages={237--248},
  year={2013}
}

@article{roloff1987interpersonal,
  title={Interpersonal processes: New directions in communication research},
  author={Roloff, Michael E and Miller, Gerald R},
  year={1987},
  publisher={Sage}
}

@book{gottman2023predicts,
  title={What predicts divorce?: The relationship between marital processes and marital outcomes},
  author={Gottman, John},
  year={2023},
  publisher={Routledge}
}

@article{gao2025homework,
  title={The Homework Wars: Exploring Emotions, Behaviours, and Conflicts in Parent-Child Homework Interactions},
  author={Gao, Nan and Liu, Yibin and Tang, Xin and Liu, Yanyan and Yu, Chun and Huang, Yun and Wang, Yuntao and Salim, Flora D and Xu, Xuhai and Wei, Jun and others},
  journal={Proceedings of the ACM on Interactive, Mobile, Wearable and Ubiquitous Technologies},
  volume={9},
  number={3},
  pages={1--37},
  year={2025},
  publisher={ACM New York, NY, USA}
}

@article{pinder2018digital,
  title={Digital behaviour change interventions to break and form habits},
  author={Pinder, Charlie and Vermeulen, Jo and Cowan, Benjamin R and Beale, Russell},
  journal={ACM Transactions on Computer-Human Interaction (TOCHI)},
  volume={25},
  number={3},
  pages={1--66},
  year={2018},
  publisher={ACM New York, NY, USA}
}

@article{nilsen2012creatures,
  title={Creatures of habit: accounting for the role of habit in implementation research on clinical behaviour change},
  author={Nilsen, Per and Roback, Kerstin and Brostr{\"o}m, Anders and Ellstr{\"o}m, Per-Erik},
  journal={Implementation Science},
  volume={7},
  number={1},
  pages={53},
  year={2012},
  publisher={Springer}
}

@book{duhigg2012power,
  title={The power of habit: Why we do what we do in life and business},
  author={Duhigg, Charles},
  volume={34},
  number={10},
  year={2012},
  publisher={Random House}
}

@inproceedings{yuksel2022conversational,
  title={Conversational Agents to Support Couple Therapy},
  author={Yuksel, Berkan and Kocaballi, A Baki},
  booktitle={Proceedings of the 34th Australian Conference on Human-Computer Interaction},
  pages={291--297},
  year={2022}
}

@article{plass2019four,
  title={Four ways of considering emotion in cognitive load theory},
  author={Plass, Jan L and Kalyuga, Slava},
  journal={Educational psychology review},
  volume={31},
  number={2},
  pages={339--359},
  year={2019},
  publisher={Springer}
}

@book{Rosenberg_2015, edition={Third edition.}, title={Nonviolent communication: a language of life}, publisher={Encinitas, Calif: PuddleDancer Press}, author={Rosenberg, Marshall B.}, year={2015} }

@book{rosenberg2004we,
  title={We can work it out: Resolving conflicts peacefully and powerfully},
  author={Rosenberg, Marshall B},
  year={2004},
  publisher={PuddleDancer Press}
}

@phdthesis{little2008total,
  title={Total honesty/total heart: Fostering empathy development and conflict resolution skills. A violence prevention strategy},
  author={Little, Marion},
  year={2008}
}

@article{nvc-couple,
author = {Édua Holmström},
title = {Enhancing the effects of emotion-focused individual and couples therapy by nonviolent communication},
journal = {Person-Centered \& Experiential Psychotherapies},
volume = {22},
number = {1},
pages = {23--40},
year = {2023},
publisher = {Routledge},
doi = {10.1080/14779757.2022.2100809},
URL = {  
        https://doi.org/10.1080/14779757.2022.2100809
},
eprint = {    
        https://doi.org/10.1080/14779757.2022.2100809
}

}

@inproceedings{rehearsal,
author = {Shaikh, Omar and Chai, Valentino Emil and Gelfand, Michele and Yang, Diyi and Bernstein, Michael S.},
title = {Rehearsal: Simulating Conflict to Teach Conflict Resolution},
year = {2024},
isbn = {9798400703300},
publisher = {Association for Computing Machinery},
address = {New York, NY, USA},
url = {https://doi-org.pitt.idm.oclc.org/10.1145/3613904.3642159},
doi = {10.1145/3613904.3642159},
booktitle = {Proceedings of the 2024 CHI Conference on Human Factors in Computing Systems},
articleno = {920},
numpages = {20},
location = {Honolulu, HI, USA},
series = {CHI '24}
}

@article{pocketbot,
author = {Zheng, Qingxiao and Markazi, Daniela M. and Tang, Yiliu and Huang, Yun},
title = {"PocketBot Is Like a Knock-On-the-Door!": Designing a Chatbot to Support Long-Distance Relationships},
year = {2021},
issue_date = {October 2021},
publisher = {Association for Computing Machinery},
address = {New York, NY, USA},
volume = {5},
number = {CSCW2},
url = {https://doi-org.pitt.idm.oclc.org/10.1145/3479589},
doi = {10.1145/3479589},
journal = {Proc. ACM Hum.-Comput. Interact.},
month = oct,
articleno = {445},
numpages = {28}
}

@article{CA-couple,
  title={Non-violent communication and marital relationship: efficacy of ‘emotion-focused couples’ communication program among filipino couples},
  author={Vazhappilly, Joshy Jacob and Reyes, Marc Eric S},
  journal={Psychological Studies},
  volume={62},
  number={3},
  pages={275--283},
  year={2017},
  publisher={Springer}
}

@article{gottman2015repair,
  title={Repair during marital conflict in newlyweds: How couples move from attack--defend to collaboration},
  author={Gottman, John M and Driver, Janice and Tabares, Amber},
  journal={Journal of Family Psychotherapy},
  volume={26},
  number={2},
  pages={85--108},
  year={2015},
  publisher={Taylor \& Francis}
}

@article{whiting2016escalating,
  title={Escalating, accusing, and rationalizing: A model of distortion and interaction in couple conflict},
  author={Whiting, Jason B and Cravens, Jaclyn D},
  journal={Journal of Couple \& Relationship Therapy},
  volume={15},
  number={4},
  pages={251--273},
  year={2016},
  publisher={Taylor \& Francis}
}

@article{weingarten1987levels,
  title={Levels of marital conflict model: A guide to assessment and intervention in troubled marriages},
  author={Weingarten, Helen and Leas, Speed},
  journal={American Journal of Orthopsychiatry},
  volume={57},
  number={3},
  pages={407--417},
  year={1987},
  publisher={Wiley Online Library}
}

@article{fowers2001limits,
  title={The limits of a techinical concept of a good marriage: Exploring the role of virtue in communication skills},
  author={Fowers, Blaine J},
  journal={Journal of Marital and Family Therapy},
  volume={27},
  number={3},
  pages={327--340},
  year={2001},
  publisher={Wiley Online Library}
}

@article{fincham2003communication,
  title={Communication skills in couples: A review and discussion of emerging perspectives},
  author={Fincham, Adrian B Kelly Frank D and Beach, Steven RH},
  journal={Handbook of communication and social interaction skills},
  pages={741--770},
  year={2003},
  publisher={Routledge}
}

@article{ronan2004violent,
  title={Violent couples: coping and communication skills},
  author={Ronan, George F and Dreer, Laura E and Dollard, Katherine M and Ronan, Donna W},
  journal={Journal of family violence},
  volume={19},
  number={2},
  pages={131--137},
  year={2004},
  publisher={Springer}
}

@incollection{rabby2003computer,
  title={Computer-mediated communication effects on relationship formation and maintenance},
  author={Rabby, Michael K and Walther, Joseph B},
  booktitle={Maintaining relationships through communication},
  pages={141--162},
  year={2003},
  publisher={Routledge}
}

@article{tong12011relational,
  title={Relational maintenance and CMC},
  author={Tong$^1$, Stephanie Tom and Walther, Joseph B},
  journal={Computer-mediated communication in personal relationships},
  pages={98},
  year={2011},
  publisher={Peter Lang}
}

@article{coyne2011luv,
  title={“I luv u:)!”: A descriptive study of the media use of individuals in romantic relationships},
  author={Coyne, Sarah M and Stockdale, Laura and Busby, Dean and Iverson, Bethany and Grant, David M},
  journal={Family Relations},
  volume={60},
  number={2},
  pages={150--162},
  year={2011},
  publisher={Wiley Online Library}
}

@inproceedings{neustaedter2012intimacy,
  title={Intimacy in long-distance relationships over video chat},
  author={Neustaedter, Carman and Greenberg, Saul},
  booktitle={Proceedings of the SIGCHI conference on human factors in computing systems},
  pages={753--762},
  year={2012}
}

@article{hampton2017channels,
  title={Channels of computer-mediated communication and satisfaction in long-distance relationships},
  author={Hampton, Adam J and Rawlings, Jessica and Treger, Stanislav and Sprecher, Susan},
  journal={Interpersona: An International Journal on Personal Relationships},
  volume={11},
  number={2},
  pages={171--187},
  year={2017}
}

@article{walther2007selective,
  title={Selective self-presentation in computer-mediated communication: Hyperpersonal dimensions of technology, language, and cognition},
  author={Walther, Joseph B},
  journal={Computers in human behavior},
  volume={23},
  number={5},
  pages={2538--2557},
  year={2007},
  publisher={Elsevier}
}

@article{perry2011couples,
  title={Couples and computer-mediated communication: A closer look at the affordances and use of the channel},
  author={Perry, Martha S and Werner-Wilson, Ronald J},
  journal={Family and Consumer Sciences Research Journal},
  volume={40},
  number={2},
  pages={120--134},
  year={2011},
  publisher={Wiley Online Library}
}

@article{hancock2001impression,
  title={Impression formation in computer-mediated communication revisited: An analysis of the breadth and intensity of impressions},
  author={Hancock, Jeffrey T and Dunham, Philip J},
  journal={Communication research},
  volume={28},
  number={3},
  pages={325--347},
  year={2001},
  publisher={Sage Publications London}
}

@article{sitorus2025language,
  title={Language ambiguity and emotional barriers: Semantic and psychological approaches in interpersonal communication},
  author={Sitorus, Ari Oldwin and Lubis, Lahmuddin},
  journal={LITERACY: International Scientific Journals of Social, Education, Humanities},
  volume={4},
  number={2},
  pages={179--188},
  year={2025}
}

@article{walther1996computer,
  title={Computer-mediated communication: Impersonal, interpersonal, and hyperpersonal interaction},
  author={Walther, Joseph B},
  journal={Communication research},
  volume={23},
  number={1},
  pages={3--43},
  year={1996},
  publisher={Sage Publications London}
}

@article{kelly2012s,
  title={“It's the American lifestyle!”: An investigation of text messaging by college students},
  author={Kelly, Lynne and Keaten, James A and Becker, Bonnie and Cole, Jodi and Littleford, Lea and Rothe, Barrett},
  journal={Qualitative Research Reports in Communication},
  volume={13},
  number={1},
  pages={1--9},
  year={2012},
  publisher={Taylor \& Francis}
}

@inproceedings{chien2013whisper,
  title={The whisper pillow: a study of technology-mediated emotional expression in close relationships},
  author={Chien, Wei-Chi and Diefenbach, Sarah and Hassenzahl, Marc},
  booktitle={Proceedings of the 6th International Conference on Designing Pleasurable Products and Interfaces},
  pages={51--59},
  year={2013}
}

@inproceedings{park2013roles,
  title={The roles of touch during phone conversations: long-distance couples' use of POKE in their homes},
  author={Park, Young-Woo and Baek, Kyoung-Min and Nam, Tek-Jin},
  booktitle={Proceedings of the SIGCHI conference on human factors in computing systems},
  pages={1679--1688},
  year={2013}
}

@inproceedings{park2012couples,
  title={How do couples use CheekTouch over phone calls?},
  author={Park, Young-Woo and Bae, Seok-Hyung and Nam, Tek-Jin},
  booktitle={Proceedings of the SIGCHI Conference on human factors in computing systems},
  pages={763--766},
  year={2012}
}

@inproceedings{griggio2019augmenting,
  title={Augmenting couples' communication with lifelines: Shared timelines of mixed contextual information},
  author={Griggio, Carla F and Nouwens, Midas and McGrenere, Joanna and Mackay, Wendy E},
  booktitle={Proceedings of the 2019 CHI Conference on Human Factors in Computing Systems},
  pages={1--13},
  year={2019}
}

@inproceedings{bales2011couplevibe,
  title={CoupleVIBE: mobile implicit communication to improve awareness for (long-distance) couples},
  author={Bales, Elizabeth and Li, Kevin A and Griwsold, William},
  booktitle={Proceedings of the ACM 2011 conference on Computer supported cooperative work},
  pages={65--74},
  year={2011}
}

@inproceedings{chun2025conflictlens,
  title={Conflictlens: Llm-based conflict resolution training in romantic relationship},
  author={Chun, Jiwon and Zhang, Gefei and Xia, Meng},
  booktitle={Adjunct Proceedings of the 38th Annual ACM Symposium on User Interface Software and Technology},
  pages={1--3},
  year={2025}
}

@article{khatra2024agent,
  title={Agent-based mediation on smartphone usage among co-located couples},
  author={Khatra, Karanmeet and Sin, Jaisie and Kuzminykh, Anastasia and Hasan, Khalad},
  journal={Proceedings of the ACM on Human-Computer Interaction},
  volume={8},
  number={MHCI},
  pages={1--20},
  year={2024},
  publisher={ACM New York, NY, USA}
}

@article{baughan2024supporting,
  title={Supporting hard conversations in close relationships through design},
  author={Baughan, Amanda and Tian, Larry and Shekar, Pranav and Zhang, Amy and Hiniker, Alexis},
  journal={Proceedings of the ACM on Human-Computer Interaction},
  volume={8},
  number={CSCW2},
  pages={1--22},
  year={2024},
  publisher={ACM New York, NY, USA}
}

@article{sproull1986reducing,
  title={Reducing social context cues: Electronic mail in organizational communication},
  author={Sproull, Lee and Kiesler, Sara},
  journal={Management science},
  volume={32},
  number={11},
  pages={1492--1512},
  year={1986},
  publisher={Informs}
}

@article{culnan1987information,
  title={Information technologies},
  author={Culnan, Mary J},
  journal={Handbook of Organizational Communication-An Interdisciplinary Perspective-},
  year={1987},
  publisher={SAGE Publications, Inc.}
}

@article{hancock2020ai,
  title={AI-mediated communication: Definition, research agenda, and ethical considerations},
  author={Hancock, Jeffrey T and Naaman, Mor and Levy, Karen},
  journal={Journal of Computer-Mediated Communication},
  volume={25},
  number={1},
  pages={89--100},
  year={2020},
  publisher={Oxford University Press}
}

@article{naveed2025comprehensive,
  title={A comprehensive overview of large language models},
  author={Naveed, Humza and Khan, Asad Ullah and Qiu, Shi and Saqib, Muhammad and Anwar, Saeed and Usman, Muhammad and Akhtar, Naveed and Barnes, Nick and Mian, Ajmal},
  journal={ACM Transactions on Intelligent Systems and Technology},
  volume={16},
  number={5},
  pages={1--72},
  year={2025},
  publisher={ACM New York, NY}
}

@article{siddals2024happened,
  title={“It happened to be the perfect thing”: experiences of generative AI chatbots for mental health},
  author={Siddals, Steven and Torous, John and Coxon, Astrid},
  journal={Npj mental health research},
  volume={3},
  number={1},
  pages={48},
  year={2024},
  publisher={Nature Publishing Group UK London}
}

@article{lee2020designing,
  title={Designing a chatbot as a mediator for promoting deep self-disclosure to a real mental health professional},
  author={Lee, Yi-Chieh and Yamashita, Naomi and Huang, Yun},
  journal={Proceedings of the ACM on Human-Computer Interaction},
  volume={4},
  number={CSCW1},
  pages={1--27},
  year={2020},
  publisher={ACM New York, NY, USA}
}

@inproceedings{zheng2025customizing,
  title={Customizing emotional support: How do individuals construct and interact with LLM-powered chatbots},
  author={Zheng, Xi and Li, Zhuoyang and Gui, Xinning and Luo, Yuhan},
  booktitle={Proceedings of the 2025 CHI conference on human factors in computing systems},
  pages={1--20},
  year={2025}
}

@inproceedings{wolfe2025toward,
  title={Toward Nonviolent Design: Co-Designing a Human-Centered Framework for AI-Mediated Communication},
  author={Wolfe, Robert and Dangol, Aayushi and Kim, JaeWon and Hiniker, Alexis},
  booktitle={Proceedings of the AAAI/ACM Conference on AI, Ethics, and Society},
  volume={8},
  number={3},
  pages={2718--2729},
  year={2025}
}

@inproceedings{fu2024text,
  title={From text to self: Users’ perception of AIMC tools on interpersonal communication and self},
  author={Fu, Yue and Foell, Sami and Xu, Xuhai and Hiniker, Alexis},
  booktitle={Proceedings of the 2024 CHI Conference on Human Factors in Computing Systems},
  pages={1--17},
  year={2024}
}

@article{erickson1993reconceptualizing,
  title={Reconceptualizing family work: The effect of emotion work on perceptions of marital quality},
  author={Erickson, Rebecca J},
  journal={Journal of Marriage and the Family},
  pages={888--900},
  year={1993},
  publisher={JSTOR}
}

@article{holm2001association,
  title={The association between emotion work balance and relationship satisfaction of couples seeking therapy},
  author={Holm, Kristen E and Werner-Wilson, Ronald J and Cook, Alicia S and Berger, Peggy S},
  journal={American Journal of Family Therapy},
  volume={29},
  number={3},
  pages={193--205},
  year={2001},
  publisher={Taylor \& Francis}
}

@article{erickson2005emotion,
  title={Why emotion work matters: Sex, gender, and the division of household labor},
  author={Erickson, Rebecca J},
  journal={Journal of marriage and family},
  volume={67},
  number={2},
  pages={337--351},
  year={2005},
  publisher={Wiley Online Library}
}

@article{gross2002emotion,
  title={Emotion regulation: Affective, cognitive, and social consequences},
  author={Gross, James J},
  journal={Psychophysiology},
  volume={39},
  number={3},
  pages={281--291},
  year={2002},
  publisher={Wiley Online Library}
}

@inproceedings{cox2016friction,
author = {Cox, Anna L. and Gould, Sandy J.J. and Cecchinato, Marta E. and Iacovides, Ioanna and Renfree, Ian},
title = {Design Frictions for Mindful Interactions: The Case for Microboundaries},
year = {2016},
isbn = {9781450340823},
publisher = {Association for Computing Machinery},
address = {New York, NY, USA},
url = {https://doi-org.proxy.lib.umich.edu/10.1145/2851581.2892410},
doi = {10.1145/2851581.2892410},
booktitle = {Proceedings of the 2016 CHI Conference Extended Abstracts on Human Factors in Computing Systems},
pages = {1389–1397},
numpages = {9},
location = {San Jose, California, USA},
series = {CHI EA '16}
}

@incollection{ryan2024self,
  title={Self-determination theory},
  author={Ryan, Richard M and Deci, Edward L},
  booktitle={Encyclopedia of quality of life and well-being research},
  pages={6229--6235},
  year={2024},
  publisher={Springer}
}

@article{vansteenkiste2020basic,
  title={Basic psychological need theory: Advancements, critical themes, and future directions},
  author={Vansteenkiste, Maarten and Ryan, Richard M and Soenens, Bart},
  journal={Motivation and emotion},
  volume={44},
  number={1},
  pages={1--31},
  year={2020},
  publisher={Springer}
}

\appendix
\newpage

\section{System Prompts}
\label{apd:prompt}

\subsection{Basic Reminder}
\begin{promptbox}{NVC-Prompt}
Your task is to analyze and detect the input from user \{user\_id\} to \{partner\_id or 'the other person'\} based on the theory of Nonviolent Communication (NVC). Strictly adhere to the ``minimum necessary intervention'' principle.

First, you need to internally determine which category the user's input belongs to (no need to output the judgment label):
\begin{itemize}
    \item \textbf{Destructive Communication Requiring Guidance}: The core characteristic of this type of language is that it includes direct personal judgments, labeling, commands, or accusations aimed at \{partner\_id or 'the other person'\} \textit{(e.g., ``You are just selfish,'' ``You are useless,'' ``You must listen to me'')}, or it contains very obvious sarcasm and taunts intended to hurt. This type of communication genuinely hinders understanding and connection, substantially damaging the relationship or \{partner\_id or 'the other person'\}.
    
    \item \textbf{Communication within Normal Range}: This is the vast majority of cases. It includes, but is not limited to: greetings, being affectionate or coy, flirting, normal statements, requests for communication, and sharing personal negative emotions (but not personal attacks). \textit{For example, {"Hey, I'm not going"}, {"Babe, don't be mad, it's my fault"}, {"I'm not going away, I just want to stick with you"}, {"I'm so tired and feel down"}, or {"You'll be the death of me one day, hmph!"}} are all considered common and acceptable interactions in an intimate relationship and thus do not require intervention.
\end{itemize}

Execute the corresponding action based on your judgment:
\begin{itemize}
    \item \textbf{If judged as ``Destructive Communication Requiring Guidance''}, your sole output must be and only be the three words \textbf{`Violent language detected'}, without any other content.
    
    \item \textbf{If judged as ``Communication within Normal Range''}, your sole output must be and only be the four words \textbf{`No violent language included'}, without any other content.
\end{itemize}
\end{promptbox}

\subsection{Neutral Guide}
\begin{promptbox}{NVC-Prompt}
Your task is to analyze and guide the input from user \{user\_id\} to \{partner\_id or 'the other person'\} based on the theory of Nonviolent Communication (NVC). Strictly adhere to the ``minimum necessary intervention'' principle.

First, you need to internally determine which category the user's input belongs to (no need to output the judgment label):
\begin{itemize}
    \item \textbf{Destructive Communication Requiring Guidance}: The core characteristic of this type of language is that it includes direct personal judgments, labeling, commands, or accusations aimed at \{partner\_id or 'the other person'\} \textit{(e.g., ``You are just selfish,'' ``You are useless,'' ``You must listen to me'')}, or it contains very obvious sarcasm and taunts intended to hurt.
    
    \item \textbf{Communication within Normal Range}: This is the vast majority of cases. It includes, but is not limited to: greetings, flirting, and sharing personal negative emotions (but not personal attacks). \textit{For example, {"Hey, I'm not going"}, {"Babe, don't be mad, it's my fault"}, {"I'm not going away, I just want to stick with you"}, {"I'm so tired and feel down"}, or {"You'll be the death of me one day, hmph!"}} are considered acceptable interactions.
\end{itemize}

Execute the corresponding action based on your judgment:
\begin{itemize}
    \item \textbf{If judged as ``Communication within Normal Range''}, your sole output must be and only be the four words \textbf{`No violent language included'}, without any other content.
    
    \item \textbf{If judged as ``Destructive Communication Requiring Guidance''}, your response must be a well-structured text with natural paragraphing and a professional tone, and it must absolutely not contain any list numbering (like a, b, c, d), titles, or stage-descriptive words. The response must naturally incorporate the following four core parts:
    \begin{itemize}
        \item \textbf{The beginning of the response} should, with an empathetic tone, analyze the possible emotions and unmet needs of user \{user\_id\}.
        \item \textbf{Next}, gently but precisely point out the specific content in the original message that does not align with NVC principles.
        \item \textbf{Subsequently}, naturally integrate a brief summary of the four elements of Nonviolent Communication (Observation, Feelings, Needs, Requests).
        \item \textbf{Finally}, provide one specific, polished example sentence as a constructive communication suggestion.
    \end{itemize}
\end{itemize}
All your outputs must strictly follow the logic described above and be written in a professional, concise, and empathetic tone.
\end{promptbox}

\begin{promptbox}{NVC-Guide}
\textbf{Your task is}: Based only on the most recent interaction and topic in the chat content provided by the user, quote statements from the chat log to help the user empathize with their partner's \textit{``emotions,''} understand their partner's \textit{``feelings,''} and guide the user to analyze and uncover all of their partner's \textit{``needs''} as comprehensively and carefully as possible, to help them understand each other better.

\textbf{Specifically}, you must remind and guide the user that the core of Nonviolent Communication is understanding one's own and the other's feelings and needs. \textbf{Consequently}, guide the user to persistently try to understand the true needs behind their own and their partner's emotions.

The user is named \{user\_id\}, their partner is named \{partner\_id\}, and the partner's gender is \{partner\_gender\}.

\textbf{Analyze the last 20 user messages. Keep the response under 400 characters.}
\end{promptbox}

\subsection{Empathetic Guide}
\begin{promptbox}{NVC-Prompt}
You are a friend and guide who is particularly skilled in handling intimate relationships using Nonviolent Communication (NVC). Your speaking style must always be gentle, approachable, and natural.

Your task is like being a secret `emotion radar'. You need to determine whether a gentle reminder is needed based on what \{user\_id\} says to \{partner\_id or 'the other person'\}. Remember, our principle is ``minimum necessary intervention''.

There are only two situations you need to make a judgment on:

\begin{itemize}
    \item \textbf{Words needing gentle guidance}: This includes direct personal judgments, labeling, commands, or accusations aimed at \{partner\_id or 'the other person'\} \textit{(e.g., ``You are just selfish,'' ``You are useless'')}.
    
    \item \textbf{Normal communication not requiring intervention}: This is the vast majority of cases! It includes being affectionate, flirting, joking, and expressing one's own grievances and emotions without personal attacks. \textit{For example, ``Hey, I'm not going'', ``Babe, don't be mad, it's my fault'', ``I'm not going away, I just want to stick with you'', ``I'm so tired and feel down'', or ``You'll be the death of me one day, hmph!''} These are healthy and normal interactions.
\end{itemize}

Now, help them based on your judgment\textasciitilde
\begin{itemize}
    \item \textbf{If you detect ``Words needing gentle guidance''}, then it's your time to shine:
    \begin{itemize}
        \item \textbf{First}, try to understand the possible emotions and thoughts of \{user\_id\}.
        \item \textbf{Then}, gently but precisely point out which expressions don't align with NVC.
        \item \textbf{Next}, use a single sentence to naturally remind them of the four key points of NVC.
        \item \textbf{Finally}, help \{user\_id\} polish their phrasing.
    \end{itemize}
    
    \item \textbf{And if you detect ``Normal communication not requiring intervention''}, your task is super simple:
    \begin{itemize}
        \item Your reply must be and can only be the four words \textbf{`No violent language included'}. Not one word more, and don't add any emojis! That's our little agreement.
    \end{itemize}
\end{itemize}
Please remember, the entire response should have natural paragraphing, without adding titles or internal classification labels.
\end{promptbox}

\begin{promptbox}{NVC-Guide}
\textbf{You are a Nonviolent Communication expert} specializing in intimate relationships, skilled at analyzing feelings, emotions, and needs. \textbf{Your tone in the response must always be}: professional yet concise, gentle and approachable, highly empathetic, persuasive, and guiding. Make the user feel cared for and supported, not criticized or lectured.

\textbf{Based only on the most recent interaction and topic} in the chat content provided by the user, quote statements from the chat log to help the user empathize with their partner's \textit{``emotions,''} understand their partner's \textit{``feelings,''} and guide the user to analyze and uncover all of their partner's \textit{``needs''} as comprehensively and carefully as possible, to help them understand each other better.

\textbf{Specifically}, you must remind and guide the user that the core of Nonviolent Communication is understanding one's own and the other's feelings and needs. \textbf{Consequently}, guide the user to persistently try to understand the true needs behind their own and their partner's emotions.

The user is named \{user\_id\}, their partner is named \{partner\_id\}, and the partner's gender is \{partner\_gender\}.

\textbf{Analyze the last 20 user messages. Keep the response under 400 characters, ensure it is clearly structured.}
\end{promptbox}

\section{Potential Conflict Topics in Romantic Relationships}
\label{apd:topics}
The following list of potential conflict topics was developed and categorized with reference to \citet{meyer2022relationship}.

\begin{enumerate}
    \item \textbf{Communication and Interaction}
        \begin{itemize}
            \item Communication styles
            \item Quality time
            \item Screen time
            \item Entertainment choices
        \end{itemize}
    \item \textbf{Personal Habits and Roles}
        \begin{itemize}
            \item Personal or partner habits
            \item Gender roles/dilemmas
            \item Role expectations
        \end{itemize}
    \item \textbf{Finances and Decision-Making}
        \begin{itemize}
            \item Spending habits
            \item Decision-making
            \item Financial management
        \end{itemize}
    \item \textbf{Future Planning}
        \begin{itemize}
            \item Future residence
            \item Career planning/work expectations
        \end{itemize}
    \item \textbf{Family Responsibilities and Lifestyle}
        \begin{itemize}
            \item Household chores distribution
            \item Daily routines and time management
            \item Food/meal choices
            \item Pet care
        \end{itemize}
    \item \textbf{Family and Social Relationships}
        \begin{itemize}
            \item Family values and conflicts
            \item Relationship with in-laws
            \item Parenting
            \item Religious and political views
        \end{itemize}
    \item \textbf{Intimate Relationships}
        \begin{itemize}
            \item Sex life
            \item Past partners
        \end{itemize}
\end{enumerate}

\section{Post-Session Questionnaire}
\label{apd:questionnaire}

All questionnaire items were assessed using a 7-point Likert scale. To evaluate the interventions, we categorized the items into four primary constructs. Unless otherwise specified in parentheses, the scale anchors are \textit{1 for strongly disagree} and \textit{7 for strongly agree}.

\vspace{2mm}
\noindent\textbf{Construct 1: System Acceptance and Perceived Effectiveness}
\begin{itemize}
    \item[\textbf{Q1.}] Overall effectiveness \textit{(1 for not effective at all, 7 for very effective)}
    \item[\textbf{Q2.}] Acceptance (refers to whether you found the intervention helpful, whether you are willing to continue participating, or whether you would recommend it to others.) \textit{(1 for not at all acceptable, 7 for very acceptable)}
\end{itemize}

\vspace{2mm}
\noindent\textbf{Construct 2: Behavioral and Interaction Changes}
\begin{itemize}
    \item[\textbf{Q4.}] The degree of my use of verbal aggression decreased.
    \item[\textbf{Q5.}] The frequency of my use of verbal aggression decreased.
    \item[\textbf{Q6.}] The degree of my partner's use of verbal aggression decreased.
    \item[\textbf{Q7.}] The frequency of my partner's use of verbal aggression decreased.

    \item[\textbf{Q3.}] Compared to previous conflicts on similar topics/situations, how has the effectiveness of communication between both parties changed? \textit{(1 for no improvement at all, 7 for significant improvement)}
    \item[\textbf{Q13.}] My ability to better clearly communicate what I need from my partner has improved.
    \item[\textbf{Q15.}] The balance between what I give and receive when communicating. \textit{(1 for less satisfied, 7 for more satisfied)}
\end{itemize}

\vspace{2mm}
\noindent\textbf{Construct 3: Cognitive and Emotional Shifts}
\begin{itemize}
    \item[\textbf{Q8.}] I more frequently examine my feelings and emotions.
    \item[\textbf{Q9.}] I am less likely to become emotional during conflicts.
    \item[\textbf{Q12.}] I have a clearer idea about why I've behaved in a certain way.

    \item[\textbf{Q10.}] My partner is less likely to become emotional or defensive during conflicts.
    \item[\textbf{Q11.}] In my relationship, I believe that there are two sides to every question, and I try to look at them both more.
    \item[\textbf{Q14.}] My partner has made more effort to understand my point of view.
\end{itemize}

\vspace{2mm}
\noindent\textbf{Construct 4: Perceived Change in Relational Quality}
\begin{itemize}
    \item[\textbf{Q16.}] After your discussion today, did your relationship become...? \textit{(1 for more distant, 7 for closer)}
    \item[\textbf{Q17.}] After your discussion today, did your relationship become...? \textit{(1 for weaker, 7 for stronger)}
    \item[\textbf{Q18.}] After your discussion today, did your relationship become...? \textit{(1 for more sad, 7 for happier)}
\end{itemize}

\vspace{2mm}
\noindent\textbf{Qualitative Feedback}
\begin{itemize}
    \item[\textbf{Q19.}] Please provide a detailed description of your comment and experience of today's intervention. \textit{(Open-ended text)}
\end{itemize}

\end{document}